\documentclass[aps, preprint]{revtex4-2}

\usepackage[utf8]{inputenc}
\usepackage[T1]{fontenc}
\usepackage{amsmath}
\usepackage{amsthm}
\usepackage{amssymb}
\usepackage{hyperref}
\usepackage{dsfont}
\usepackage{graphicx}
\usepackage{enumerate}
\usepackage{bm}

\hypersetup{colorlinks=true, linkcolor=blue, citecolor=blue}

\newcommand{\Real}{\mathbb{R}}

\newcommand{\Comp}{\mathbb{C}}

\newcommand{\pspace}{\Real^3 \setminus \{0\}}

\newcommand{\Lightcone}{\mathcal{L}_+}

\newcommand{\Poincare}{Poincar\'{e} }
\newcommand{\SO}{\mathrm{SO}}
\newcommand{\so}{\mathfrak{so}}

\newcommand{\ISO}{\mathrm{ISO}}

\newcommand{\re}{\text{Re}}

\newcommand{\uv}[1]{\mbf{#1}}
\newcommand{\mbf}[1]{\boldsymbol{#1}}
\newcommand{\boldcal}[1]{\mbf{\mathcal{#1}}}
\newcommand{\kvec}{\mbf{k}} 

\newcommand{\khat}{\hat{\kvec}}
\newcommand{\sign}{\text{sgn}}
\newcommand{\etheta}{\uv{e}_\theta}
\newcommand{\ephi}{\uv{e}_\phi}

\newcommand{\eh}{\uv{e}_h}
\newcommand{\epm}{\uv{e}_\pm}

\newcommand{\phat}{\hat{\mbf{P}}}

\newtheorem{prop}{Proposition}

\begin{document}

\title{The topology, geometry, and angular momentum of unmagnetized plasma waves}
\date{\today}

\author{Eric Palmerduca}
\email{ep11@princeton.edu}
\affiliation{Department of Astrophysical Sciences, Princeton University, Princeton, NJ 08544, USA}
\affiliation{Princeton Plasma Physics Laboratory, Princeton, NJ 08543,
USA}

\author{Hong Qin}
\email{hongqin@princeton.edu}
\affiliation{Department of Astrophysical Sciences, Princeton University, Princeton, NJ 08544, USA}
\affiliation{Princeton Plasma Physics Laboratory, Princeton, NJ 08543,
USA}

\begin{abstract}
It was recently discovered that plasma waves possess topologically protected edge modes, indicating the existence of topologically nontrivial structures in the governing equations. Here we give a rigorous study of the underlying topological vector bundle structure of cold unmagnetized plasma waves and show that this topology can be used to uncover a number of new results about these waves. The topological properties of the electromagnetic waves mirror those recently found for photons and other massless particles. We show that there exists an explicit globally smooth polarization basis for electromagnetic plasma waves---surprisingly, this does not violate the hairy ball theorem. The rotational symmetry of the waves gives a natural decomposition into topologically nontrivial $R$ and $L$ circularly polarized electromagnetic waves and the topologically trivial electrostatic Langmuir waves. The existence of topologically nontrivial waves, despite the effective mass introduced by the plasma, is related to the resonance of electrostatic and electromagnetic waves. We show that the eigenstates of the angular momentum operator are the spin-weighted spherical harmonics, giving a novel globally smooth basis for plasma waves. The sparseness of the resultant angular momentum multiplet structure illustrates that the angular momentum does not split into well-defined spin and orbital parts.  However, we demonstrate that the angular momentum admits a natural decomposition, induced by the rotational symmetry,  into two quasi-angular momentum components, termed helicity and orbital quasi-angular momentum. Although these operators do not generate physical rotations and therefore do not qualify as true angular momentum operators, they are gauge invariant, well-defined, and appear to be experimentally relevant. 
\end{abstract}

\maketitle

\section{Introduction}
Topology has been at the root of a number of major discoveries in the physics of lattice media, including the quantum Hall effects \cite{Thouless1982, Avron1983,Kohmoto1985}, the intrinsic anomalous Hall effect \cite{Haldane2004}, the bulk-edge correspondence (BEC) \cite{Schulz2000,Bernevig2013}, and the existence of topological insulators \cite{Kane2005,Hasan2010}. It has been recently found that waves in continuous media such as fluids \cite{Souslov2017,Delplace2017,Tauber2019,Souslov2019,Perrot2019,Faure2023} and plasmas \cite{Yang2016,Gao2016,Parker2020a,Fu2021,Fu2022,Qin2023,Qin2024plasma} can also exhibit topological behavior. So far, topological applications in continuous media have mostly focused on the use of the BEC to predict the existence of topologically protected edge modes. In this method, one typically follows a prescription of calculating the Berry connection and Berry curvature, and the latter is then integrated over some 2D surface in $k$ space to obtain a topological Chern number. The difference in Chern numbers across a boundary then determines the number of edge modes via the BEC. Although it is rarely mentioned in the waves literature, Chern numbers are associated with vector bundles \cite{Tu2017differential, Faure2023, Qin2023}, which are parameterized collections of vector spaces. The success of the BEC in plasma physics indicates that plasma waves possess such a topological structure, in particular, the collection of all plane wave solutions forms a vector bundle parameterized by the momentum space \cite{Qin2023}. It was recently shown in the context of massless particle waves, such as photons and gravitons, that the vector bundle structure can be used to explore topological properties of waves besides just the BEC \cite{PalmerducaQin_PT, PalmerducaQin_GT, PalmerducaQin_helicity, PalmerducaQin_SAMOAM,PalmerducaQin_SWSH}. In this article, we show that similar techniques can be used to study the topology of waves in cold unmagnetized plasmas, and in fact, electromagnetic (EM) plasma waves have essentially the same topological structure as photons. We will show that there exists a global polarization basis for EM plasma waves, which is surprising as one might guess that this would violate the hairy ball theorem. We further show that any such basis must involve elliptical polarizations as it is not possible to form a global basis using only circularly or linearly polarized waves. 

Our main use of the vector bundle structure pertains to the angular momentum of plasma waves and the question of whether or not it permits a spin-orbital decomposition. In laser-plasma accelerators, the angular momentum of the electron beam is one of the two key parameters (along with the emittance) for characterizing the beam profile \cite{Phuoc2006,Thaury2013}. Recent simulations have suggested that the use of so-called orbital angular momentum (OAM) beams in such accelerators may lead to enhanced tuning of the electron beam and the resultant X-ray production \cite{Martins2019,Shi2021}. It has also been suggested that OAM beams may be used in the production of extremely large magnetic fields in plasmas, with simulations suggesting that fields in the kilo-Tesla range may be accessible for short pulses \cite{Longman2021,Longman2022} and $\sim40 \text{ T}$ in the steady state \cite{Ji2023}. Despite the great success of so-called OAM beams in applications \cite{Shen2019}, it has long been noted that there are deep theoretical issues with the concept of splitting the angular momentum of light waves into spin angular momentum (SAM) and OAM, and that the apparent splitting in Laguerre-Gaussian modes does not hold beyond the paraxial approximation \cite{VanEnk1994_EPL_1,Barnett1994,Allen1999}. The fundamental issue is that the alleged SAM and OAM operators for light do not satisfy angular momentum commutation relations and are thus not true angular momentum operators \cite{VanEnk1994_EPL_1,VanEnk1994_JMO_2,Leader2019, PalmerducaQin_PT, PalmerducaQin_SAMOAM}. We show here that the same is true for cold plasma waves. Nevertheless, we show that the rotational symmetry of plasma waves induces a unique splitting of the angular momentum operator $\boldsymbol{J} = \boldsymbol{J}_{\parallel} + \boldsymbol{J}_\perp$, where $\boldsymbol{J}_\parallel$ and $\boldsymbol{J}_\perp$ are well-defined operators. We term these the helicity quasi-angular momentum and orbital quasi-angular momentum, respectively. They are not true angular momentum operators since they do not satisfy angular momentum commutation relations (\emph{i.e.}, they do not generate $\SO(3)$ symmetries). We show, for example, that the components of $\mbf{J}_\parallel$ and $\mbf{J}_\perp$ have continuous spectra, something which is impossible for legitimate angular momentum operators. Despite this deficiency, these operators appear to be of both theoretical and experimental interest. 

We also use the angular momentum operator $\mbf{J}$, which is the generator of the $\SO(3)$ symmetry of the plasma waves, to give a novel harmonic decomposition of these waves. Indeed, plasma waves can be decomposed into rotationally invariant finite-dimensional multiplets of wave solutions. More particularly, these are simultaneous eigenstates of energy, helicity, $J^2$, and $J_z$. This gives a basis for plasma waves in terms of spin-weighted spherical harmonics, which are generalizations of the more common ordinary spherical harmonics. The resultant angular momentum multiplet structure gives a concrete illustration of the issues that arise when attempting a spin-orbital decomposition for plasma waves.

This paper is organized as follows. In Sec. \ref{sec:plasma:Vector_bundle_structure}, we describe the vector bundle formalism for plasma waves. In Sec. \ref{sec:plasma:topology_of_plasma_waves}, we characterize the topology of the EM and electrostatic (ES) Langmuir waves, showing that Langmuir waves are topologically trivial, as is the collection of all EM plasma waves. This allows for an explicit construction of global polarization bases for EM and ES waves which vary smoothly with the momentum $\kvec$. However, if we decompose the EM waves into rotationally invariant parts, namely $R$ and $L$ waves, then we find that these subspaces are topologically nontrivial. Although the plasma introduces an effective mass, which typically indicates topological triviality, we show that topologically nontrivial waves are able to exist due to the resonance of electromagnetic and electrostatic waves at the plasma frequency. In Sec. \ref{sec:plasma:AM_of_plasma_waves}, we show that we can define a global basis for $R$, $L$, and Langmuir waves in terms of SWSHs, and how this harmonic decomposition illustrates the obstruction to splitting the angular momentum of cold plasma waves into SAM and OAM parts. In Sec. \ref{sec:plasma:plasma_par_perp_decomp}, we show that there is a symmetry-induced decomposition of the plasma angular momentum, but that it is not a legitimate SAM-OAM splitting. Sec. \ref{sec:plasma:kinetic} discusses the generalization of these results to kinetic plasmas.

\section{Vector bundle structure of cold plasma waves}\label{sec:plasma:Vector_bundle_structure}
We will consider linear waves in a cold homogeneous unmagnetized plasma. Assume that electrons have density $n = n_0 + n_1$ where $n_0$ is constant and $n_1 \ll n_0$. We will work in the background plasma's rest frame, and let $\mbf{v}$ be the perturbed electron velocity. With ions treated as a neutralizing background, the linearized fluid equations are
\begin{gather}
    m\frac{\partial \mbf{v}}{\partial t} = -e\boldcal{E} \\
    \frac{\partial n_1}{\partial t} = -n_0 \nabla \cdot \mbf{v} \\
    \frac{\partial \boldcal{E}}{\partial t} = c \nabla \times \boldcal{B}  + 4\pi e n_0\mbf{v}\\
    \frac{\partial \boldcal{B}}{\partial t} = -c \nabla \times \boldcal{E} \\
    \nabla \cdot \boldcal{E} = -4\pi e n_1 \\
    \nabla \cdot \boldcal{B} = 0
\end{gather}
where $e>0$ and $m$ is the electron mass. We consider plane wave solutions with $(t,\mbf{x})$ dependence of the form $e^{i(\mbf{k}\cdot \mbf{x}-\omega t)}$. The system simplifies to an eigenvalue problem
\begin{gather}\label{eq:fluid_dispersion_tensor}
    \mbf{D}_E(\omega,\kvec)\cdot \mbf{E}(\mbf{k}) = 0 \\
    D_{E,ab}(\omega, \kvec) \doteq \frac{1}{\omega^2}\big[k_a k_b  + \delta_{ab} (\omega^2 - \omega_{p}^2 - k^2)\big]
\end{gather}
where $D_{E,ab}$ is the dispersion tensor and $\omega_{p} = (4\pi n_0e^2 / m)^{1/2}$ is the background plasma frequency. $\boldcal{E}(\mbf{x},t)$ and $\mbf{E}(\mbf{k})$ are related by
\begin{equation}
    \boldcal{E}(\mbf{x},t) = \frac{1}{2(2\pi)^{3/2}}\int d^3k \, \mbf{E}(\kvec)e^{i(\mbf{k}\cdot\mbf{x} - \omega(\kvec) t)} + c.c..
\end{equation}
The other dynamical variables are determined by $\mbf{E}(\kvec)$:
\begin{gather}
    n_1(\kvec) = -\frac{i}{4\pi e}\kvec \cdot \mbf{E}(\kvec) \\
        \mbf{v}(\kvec) = -\frac{ie}{m\omega}\mbf{E}(\kvec) \\
        \mbf{B}(\kvec) = \frac{c}{\omega}\kvec \times \mbf{E}(\kvec).
\end{gather}
Note that we use the same symbols $n_1$ and $\mbf{v}$ for the position space variables and their Fourier transform; it will always be clear from context which is being used. The branches of the solution are determined by the dispersion relation $\det \mbf{D} = 0$. The positive frequency solutions correspond to a nondegenerate electrostatic (ES) Langmuir mode with $\omega = \omega_{p}$ and two degenerate EM modes with $\omega = \sqrt{\omega_{p}^2 + |\kvec|^2c^2}$ \footnote{Modes can be unambiguously labeled as ES or EM when working in the background plasma rest frame. We note, however, that this distinction is not valid in reference frames where the background plasma is flowing \cite{Qin2024plasma}.}. The ES modes can be labeled by $(k,E\khat)$ where $E \in \Comp$ and $k^\mu = (\omega_{p}/c,\kvec)$ is the wave four-vector. We denote the collection of all such modes by $\zeta_0$. For each fixed four-vector $k_0$, the subset $\zeta_0(k_0)$ of all modes with momentum $k_0$ comprise a one-dimensional vector space. Similarly, the EM modes can be labeled by $(k,\mbf{E})$ for $k = (\sqrt{\omega_{p}^2/c^2 + |\kvec|^2},\kvec)$ and any $\mbf{E} \in \Comp^3$ satisfying the transversality condition $\mbf{E} \cdot \kvec = 0$. Denote the collection of all such modes by $\zeta_{EM}$ and those with momentum $k_0$ by $\zeta_{EM}(k_0)$. Each $\zeta_{EM}(k_0)$ is a two-dimensional vector space. Thus, $\zeta_0$ and $\zeta_{EM}$ form collections of vector spaces smoothly parameterized by subsets of the four-momentum space; that is, $\zeta_0$ and $\zeta_{EM}$ are vector bundles \cite{Tu2017differential}. The parameterizing space is known as the base manifold. Note that at $(\omega/c,\kvec) = (\omega_{p}/c,0)$, all three modes degenerate and it is not possible to identify subspaces corresponding to ES and EM modes. In other words, the limits $\lim_{\kvec\rightarrow 0} \zeta_0(k)$ and $\lim_{\kvec\rightarrow 0} \zeta_{EM}(k)$ do not exist as vector spaces. It is thus necessary to remove the degeneracy point $(\omega/c,\kvec) = (\omega_p/c,\mbf{0})$ from the base manifold. For EM modes, the base manifold is the mass hyperboloid
\begin{equation}
    \mathcal{M}_{\omega_{p}} = \Big\{(\omega/c,\kvec) \in \Real^{3+1}\setminus\{(\omega_p/c,\mbf{0})\} | \omega = \sqrt{\omega_{p}^2 + |\kvec|^2c^2} \Big\}.
\end{equation}
\begin{figure}
    \centering
    \includegraphics[width=8.6cm]{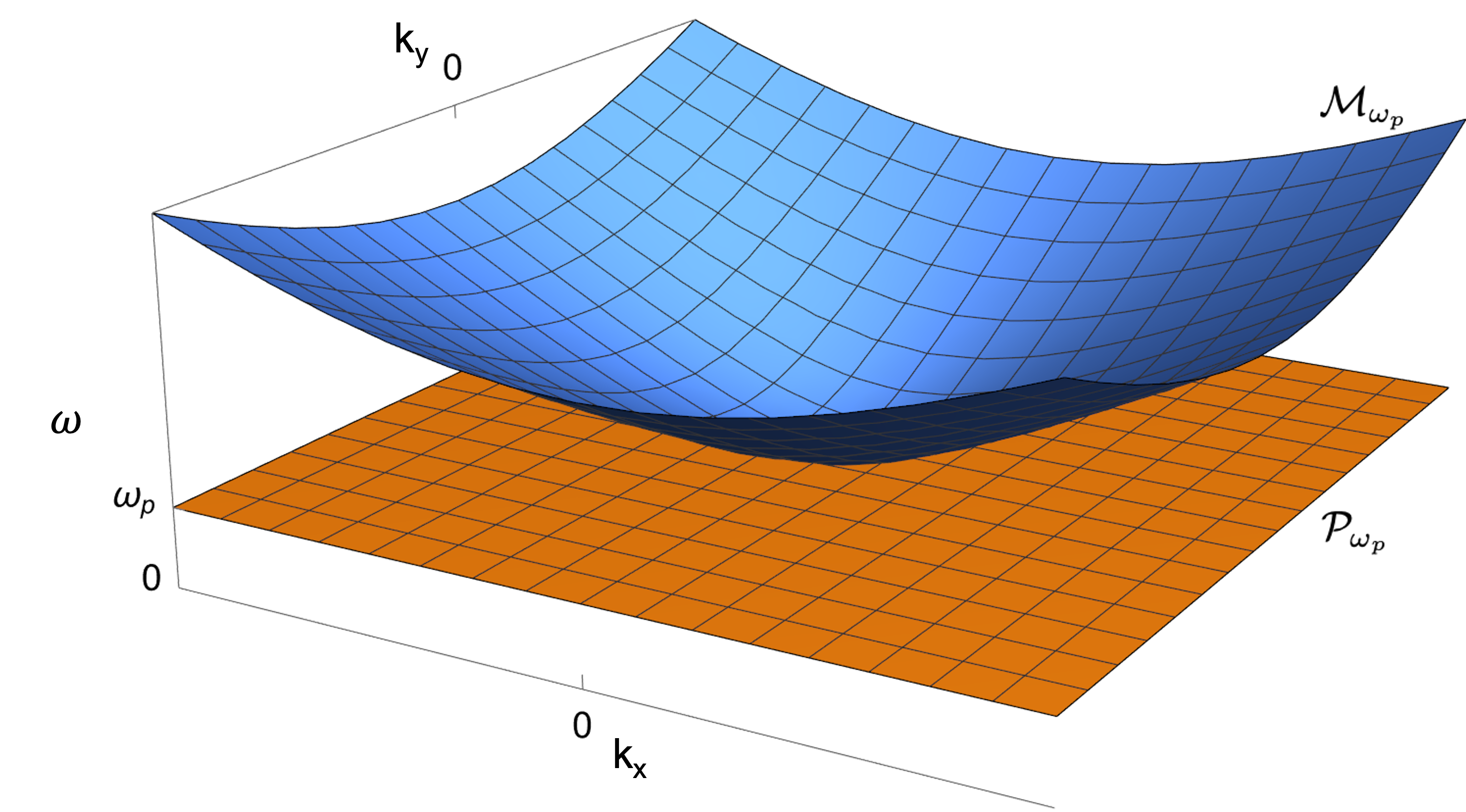}
    \caption{Electrostatic (orange) and electromagnetic (blue) dispersion relations for fixed $k_z = 0$. The electrostatic waves reside on the plane $\mathcal{P}_{\omega_p}$ in four-momentum space, while electromagnetic waves reside on a mass hyperboloid $\mathcal{M}_{\omega_p}$. The waves degenerate at $\mbf{k} = 0$, $\omega = \omega_p$.}\label{fig:plasma_dispersion}
\end{figure}%
For ES modes, the base manifold is the plane $\omega = \omega_{p}$ with the point $(\omega_{p}/c,\mbf{0})$ removed; we denote this plane by $\mathcal{P}_{\omega_{p}}$. This is depicted in Fig. \ref{fig:plasma_dispersion}. Both $\mathcal{M}_{\omega_p}$ and $\mathcal{P}_{\omega_p}$ can be parameterized by just the three-vector $\kvec$, and so the base manifolds for ES and EM waves are homeomorphic to the punctured space $\pspace$. Using standard notation \cite{Tu_manifolds}, we write the vector bundles as $\pi_{EM}:\zeta_{EM}\rightarrow \pspace$ and $\pi_{0}:\zeta_0 \rightarrow \pspace$. Here $\pi_{EM}$ is the projection onto the base manifold $(\mbf{k},\mbf{E}) \mapsto \mbf{k}$, and similarly for $\pi_{0}$. The vector spaces $\zeta_{EM}(\kvec)$ and $\zeta_0(\kvec)$ are called the fibers at $\kvec$. As vector bundles are among the most well-studied topological objects, this description allows a direct study of the topology of plasma waves.

Vector bundles also allow for precise treatment of the geometry of waves, in particular, their metric and symmetry properties. A detailed discussion of representations of symmetry groups on vector bundles can be found in Refs. \cite{Simms1968, PalmerducaQin_PT}. Each fiber is a Hermitian vector space with inner product defined by
\begin{equation}\label{eq:plasma:Hermitian_product_bundle}
    \langle (\kvec, \mbf{E}) , (\kvec, \mbf{E}')\rangle_{\text{pw}} = \mbf{E}^* \cdot \mbf{E}',
\end{equation}
making $\zeta_{EM}$ and $\zeta_0$ into Hermitian vector bundles; the subscript indicates that this is a pointwise inner product at a specific $\kvec$. 

We now consider the spacetime symmetry of these waves. In studying the geometry of EM vacuum waves (\emph{i.e.}, photons), the fully relativistic spacetime symmetry group is relevant, namely the (proper orthochronous) \Poincare group $\ISO^+(3,1)$. This group consists of all combinations of spatial rotations, relativistic boosts, and spacetime translations. There are two differences in treating plasma waves. The most important is that the rest frame of the background plasma introduces a preferred reference frame. We have written the cold fluid equations assuming the background plasma is stationary, thus the solution space is not symmetric under boosts \footnote{Of course, we could produce a solution space that is boost symmetric by allowing the background flow velocity to be arbitrary, but this would lead to many solutions which are physically equivalent and complicate our analysis.}. The second difference is that we will work nonrelativistically, which is not possible in the case of photons. However, this does not introduce substantial modifications since this primarily affects boost transformations, which the solution space not possess anyway. Thus, the symmetry group of interest is $G = \Real^{3+1} \rtimes \SO(3)$, where $\Real^{3+1}$ corresponds to spacetime translations and $\SO(3)$ to rotations. The semidirect product $\rtimes$ reflects the fact that spatial rotations and translations do not commute, and in particular that $\mbf{a}R\mbf{a}^{-1} \notin \SO(3)$ for $\mbf{a} \in \Real^3$ and $R \in \SO(3)$. $G$ is the subgroup of the Galilean group not containing boosts; $\SO(3)$ is the homogeneous part of $G$. The actions of $a \in \Real^{3+1}$ and $R \in \SO(3)$ on a mode $(\kvec,\mbf{E})$ are described by
\begin{subequations}
\label{eq:plasma:plasma_vb_action}
\begin{gather}
    \Sigma_{a}(\kvec,\mbf{E}) = e^{-ik^\mu a_\mu}(\kvec,\mbf{E}) \\
    \Sigma_{R}(\kvec,\mbf{E}) = (R\kvec,R\mbf{E})
\end{gather}
\end{subequations}
where we use summation notation with the $(-+++)$ signature and $R\kvec$ and $R\mbf{E}$ denote the standard action of $\SO(3)$ on 3-vectors. For any $g_1,g_2 \in G$,
\begin{gather}
    \Sigma_{g_1g_2}(\kvec, \mbf{E}) = \Sigma_{g_1}\Sigma_{g_2}(\kvec,\mbf{E})
\end{gather}
so we have a vector bundle representation of $G$. The representation is unitary with respect to the Hermitian product $(\ref{eq:plasma:Hermitian_product_bundle})$:
\begin{equation}
    \langle \Sigma_g(\kvec,\mbf{E}), \Sigma_g(\kvec,\mbf{E}')\rangle_{\text{pw}} = \langle (\kvec,\mbf{E}), (\kvec,\mbf{E}')\rangle_{\text{pw}}
\end{equation}
for any $g$. 

The representation $\Sigma$ on the vector bundles also induces a representation on the Hilbert spaces of (square-integrable) sections of $\zeta_0$ and $\zeta_{EM}$, denoted by $L^2(\zeta_0)$ and $L^2(\zeta_{EM})$. A section, written as $\mbf{E}(\mbf{k})$ or $|\mbf{E}(\mbf{k})\rangle$, is a choice of vector in each fiber and describes a superposition of plane waves. The inner product of sections is given by 
\begin{equation}\label{eq:plasma:sectional_inner_product}
    \langle \mbf{E}_1(\kvec),\mbf{E}_2(\kvec)\rangle = \int \frac{d^3k'}{\omega(\kvec')} \mbf{E}^*_1(\kvec')\cdot \mbf{E}_2(\kvec'). 
\end{equation}
The action of $g \in G$ on a section $\mbf{E}(\kvec)$ induced by (\ref{eq:plasma:plasma_vb_action}) is given by
\begin{equation}
    (\tilde{\Sigma}_{g}\mbf{E})(\kvec) = \Sigma_{g}\mbf{E}\big(\Sigma_{g^{-1}}\mbf{k} \big),
\end{equation}
that is, the external degree of freedom (DOF) $\kvec$ and the internal DOF (the polarization) are simultaneously transformed. The representations of the Lie group $G = \Real^{3+1} \rtimes \SO(3)$ also induce representations $\tilde{\eta}$ of the Lie algebra $\mathfrak{g} = \Real^{3+1} \rtimes \so(3)$. In particular, if $ \mathrm{g}\in \mathfrak{g}$, then $\tilde{\eta}_\mathrm{g}$ is the operator defined by
\begin{equation}
    \tilde{\eta}_\mathrm{g} \mbf{E} = \frac{d}{dt}\Big|_{t=0}\tilde{\Sigma}_{\exp(i t\mathrm{g})}\mbf{E}.
\end{equation}
These are the generators of the symmetry $\Sigma$. The generators of the  $\SO(3)$ symmetry are the angular momentum operators $\boldsymbol{J} = (J_1,J_2,J_3)$, the generators of the $\Real^{3+1}$ translational symmetry are the momentum operators $\mbf{P} = (P_1,P_2,P_3)$ and the Hamiltonian $H$. Explicitly,
\begin{gather}
    J_n\mbf{E} = i\frac{d}{d\varphi}\Big|_{\varphi=0}\tilde{\Sigma}(R_n(\varphi))\mbf{E}, \\
    [P_n \mbf{E}](\kvec)= k_n\mbf{E}(\kvec), \\
    [H\mbf{E}](\kvec) = \omega(\kvec) \mbf{E}(\kvec),
\end{gather}
where $R_a(\varphi)$ is a rotation by angle $\varphi$ about the $n$ axis.
The expectation values of these operators coincide with the total fluid energy, momentum, and angular momentum, as shown in Sec. \ref{sec:plasma:plasma_par_perp_decomp}, Eqs. (\ref{eq:plasma:E_expectation})-(\ref{eq:plasma:J_expectation}) and Appendix \ref{appendix:A}.

\section{Topology of plasma waves}\label{sec:plasma:topology_of_plasma_waves}
Vector bundles of rank $1$ and $2$ over $\pspace$ are completely classified topologically by their (first) Chern number $C$, which can assume any integer value. This follows from noting that the topology is unaffected by deforming the base manifold, and $\pspace$ deformation retracts onto the unit sphere $S^2$. The classification via Chern numbers then follows from the corresponding classification of vector bundles over $S^2$ \cite{PalmerducaQin_PT, McDuff2017}. We can thus classify the topology of EM and ES waves via their Chern numbers. The Chern number measures how globally ``twisted'' a vector bundle is \cite{Bott1982}.

\subsection{Topology of electrostatic waves}
It is straightforward to see that ES waves are topologically trivial and thus have Chern number 0. A vector bundle is topologically trivial if and only if it has a smoothly varying basis over the whole bundle. That is to say, a rank $r$ vector bundle is trivial if it has $r$ nowhere-vanishing sections $(\psi_1,\ldots \psi_r)$ such that $(\psi_1(\kvec),\ldots, \psi_r(\kvec))$ are linearly independent every $\kvec$. Since $\zeta_0$ is a line bundle $(r=1)$, it is trivial because $\mbf{E}(\kvec) = E_0\khat$ is such a smooth basis for any nonzero $E_0 \in \Comp$. Thus, $\zeta_0$ is isomorphic to the bundle $\pspace \times \Comp$.

\subsection{Topology of \texorpdfstring{$\zeta_{EM}$}{zetaEM}}
It is much less obvious that $\zeta_{EM}$ is topologically nontrivial. Because any $(\kvec,\mbf{E}) \in \zeta_{EM}$ satisfies the transversality condition $\kvec \cdot \mbf{E} = 0$, a nonvanishing section $\mbf{E}(\kvec)$ would imply a choice of transverse vector at every point on the unit $\kvec$ sphere. Naively, this would appear to violate the hairy ball theorem (\cite{Milnor1978}, Theorem 1). However, the hairy ball theorem only applies to real vector fields \cite{PalmerducaQin_PT}; it does not apply in the present setting as $\mbf{E}(\kvec)$ is generally complex. 

Rather surprisingly, it turns out that $\zeta_{EM}$ actually is trivial. To establish this result, we will first observe that $\zeta_{EM}$ is topologically equivalent to the photon bundle $\gamma$ that was studied extensively in Ref. \cite{PalmerducaQin_PT}. The photon bundle consists of all of the vacuum solutions of Maxwell's equation in $\kvec$ space. It is constructed in exactly the same way as $\zeta_{EM}$ except with $\omega_{p} = 0$. For all values of $\omega_{p}$, the EM modes are described by $(\kvec,\mbf{E})$ with $\mbf{E}$ transverse to $\kvec$, thus, the bundles $\zeta_{EM}$ and $\gamma$ are isomorphic for all values of $\omega_{p}$, including for $\omega_{p} = 0$. Thus, we can apply all of the results from Ref. \cite{PalmerducaQin_PT} on the topology of the photon bundle to the EM plasma bundle $\zeta_{EM}$. In particular, one can show that the Chern number of $\zeta_{EM}$ is $0$ by observing that $\zeta_{EM}$ is the complexification of a real vector bundle and that the complexification of real vector bundles always have vanishing first Chern number (\cite{PalmerducaQin_PT}, Theorem 7). This is sufficient to prove that $\zeta_{EM}$ is topologically trivial, and thus there exists a globally smooth orthonormal basis for these EM waves. It takes substantially more work to explicitly find such a basis, although it can be accomplished using the clutching construction from algebraic topology \cite{HatcherVBKT}. The explicit basis found for the photon bundle in Ref. \cite{PalmerducaQin_PT} also gives a basis for $\zeta_{EM}$. Define
\begin{subequations}\begin{alignat}{2}
    \boldsymbol{\epsilon}_{1+}(\kvec) &= &&\sin(\phi) \etheta + \cos (\phi) \ephi \label{eq:plasma:v1p}\\
    \boldsymbol{\epsilon}_{2+}(\kvec) &= -&&\cos(\phi) \etheta + \sin(\phi) \ephi \\
    \boldsymbol{\epsilon}_{1-}(\kvec) &= -&&\sin(\phi) \etheta + \cos(\phi) \ephi \\
    \boldsymbol{\epsilon}_{2-}(\kvec) &= -&&\cos(\phi) \etheta - \sin(\phi) \ephi. \label{eq:plasma:v2m}
\end{alignat} \end{subequations}
where $(\theta,\phi)$ are spherical coordinates in $\kvec$ space and $(\hat{\kvec},\etheta,\ephi)$ are the standard spherical unit vectors and $\kvec \neq 0$. The orthonormal basis for $\zeta_{EM}$ is given by $[\boldsymbol{\epsilon}_1(\kvec), \boldsymbol{\epsilon}_2(\kvec)]$ where
\begin{equation}
[\boldsymbol{\epsilon}_1(\kvec), \boldsymbol{\epsilon}_2(\kvec)] = 
    \begin{cases}
        [\boldsymbol{\epsilon}_{1+}, \boldsymbol{\epsilon}_{2+}] & \text{for } 0 \leq \theta \leq \frac{\pi}{2} \\
        [\mathcal{F}(\theta,\phi)\boldsymbol{\epsilon}_{1-}, \mathcal{F}(\theta,\phi)\boldsymbol{\epsilon}_{2-}] & \text{for } \frac{\pi}{2} < \theta \leq \pi
    \end{cases}
\end{equation}
and $\mathcal{F}(\theta,\phi)$ is a 2D linear transformation which in the $(\etheta,\ephi)$ basis is given by
\begin{subequations}\begin{align}
        F(\theta,\phi) &\doteq \begin{pmatrix}
    x\big( g(\theta),\phi \big) + iz\big( g(\theta),\phi \big) && y\big( g(\theta),\phi \big) \\
    -y\big( g(\theta),\phi \big) && x\big( g(\theta),\phi \big) - iz\big(g(\theta),\phi \big)
    \end{pmatrix} \\
    x(\theta,\phi) &\doteq \cos(2\phi)\sin^2(\theta) + \cos^2(\theta), \\
    y(\theta,\phi) &\doteq \sin(2\phi)\sin(\theta), \\
    z(\theta,\phi) &\doteq -\sin^2(\phi)\sin(2\theta) \\
    h(t) &\doteq \begin{cases}
        e^{-1/t} & t > 0 \\
        0 & t \leq 0,
    \end{cases} \\
    g(\theta) &\doteq \frac{\pi}{2}\Big(1 + \frac{h(\theta - \pi/2)}{h(\theta - \pi/2) + h(\pi - \theta)} \Big).
\end{align} \end{subequations}
It is carefully constructed to be smooth at the poles and along the equator. This basis is fairly complicated, but it shows that it is possible to construct an orthonormal basis for EM waves. We suspect that alternative methods may be utilized to construct substantially simpler global bases, and we hope to explore this possibility in future work.

Since $\zeta_{EM} \cong \gamma$, all of the topological results obtained for $\gamma$ in Ref. \cite{PalmerducaQin_PT} apply directly to $\zeta_{EM}$. In particular, it is not possible to construct a global basis consisting of only linear polarizations (\cite{PalmerducaQin_PT}, Theorem 20). Intuitively, this is because linear polarizations are scalar multiples of real polarization vectors, and thus, in this case, the hairy ball theorem does apply to preclude any globally nonvanishing linearly polarized vector fields. A global linear polarization basis would imply that $\zeta_{EM} \cong \gamma \cong \ell_1 \oplus \ell_2$ where $\ell_1$ and $\ell_2$ are trivial linearly polarized line bundles. However, there do not exist \emph{any} linearly polarized subbundles of $\zeta_{EM} \cong \gamma$, let alone topologically trivial ones.

\subsection{R and L waves are topologically nontrivial}
While the complete set of EM waves is topologically trivial, it is not possible to trivialize it using only linear polarizations. One may ask an alternative question: is it possible to write a global basis with one basis vector $R$ circularly polarized and the other $L$ circularly polarized? The answer is no, but the obstruction is quite different from that which prevents a linear basis. To begin with, the circularly polarized waves become relevant once one considers the geometry of the EM waves, that is, once we consider the action of the spacetime symmetry group $G$. While $\zeta_{EM}$ is a unitary representation of $G$, it is not an irreducible representation, meaning that there are rotationally and translationally symmetric subbundles of $\zeta_{EM}$. These subbundles are the $R$ and $L$ circularly polarized subbundles $\zeta_{+}$ and $\zeta_{-}$. $\zeta_\pm$ consists of all $(\kvec,\mbf{E}) \in \zeta_{EM}$ such that $\mbf{E} = c(\uv{e}_1 \pm i\uv{e}_2)$ for some $c \in \Comp$ and some right-handed orthonormal coordinate system $(\uv{e}_1,\uv{e}_2,\khat)$. $\zeta_{EM}$ is the direct sum of these bundles
\begin{equation}
    \zeta_{EM} =  \zeta_+ \oplus \zeta_{-}.
\end{equation}
That $\zeta_{\pm}$ are irreducible representations of $G$ means that an $R$ or $L$ polarized wave will always remain $R$ or $L$ polarized after rotation or translation. It is important to note that the $R$ and $L$ waves behave differently from the spin up and spin down states of an electron or a massive spin 1 particle. The spin up and spin down states can be transformed into each other via a change in reference frame, while no change of reference frame (no element of $G$) can transform an $R$ wave into an $L$ wave. We note that decomposing a representation into its irreducible subrepresentations is a central problem in representation theory and particle physics. This is true in the latter because particles are typically defined as irreducible representations of the \Poincare group \cite{Weinberg1995,Maggiore2005}; in the present case we are essentially identifying the quasiparticles in the fluid plasma system. The Langmuir waves are also quasiparticles in this sense, as $\zeta_0$ has rank 1 and is thus irreducible. Unlike the total EM bundle $\zeta_{EM}$, the circularly polarized bundles $\zeta_{\pm}$ are not topologically trivial. This is most easily seen by calculating the Chern numbers via the Berry connection; the calculation is the same as that for vacuum photons given in Ref. \cite{PalmerducaQin_PT}, Theorem 18. One finds that the Chern numbers are $C(\zeta_\pm) = \mp 2$. The nonzero Chern numbers indicate that the bundles $\zeta_{\pm}$ are nontrivial. It is therefore not possible to form a global nonvanishing basis of either $\zeta_+$ or $\zeta_-$; put differently, if $\mbf{E}_R(\kvec)$ denotes a superposition of $R$ polarized modes which varies continuously with $\kvec$, then there must be at least one $\kvec_0$ such that $\mbf{E}_R(\kvec_0) = 0$, and similarly for $L$ polarized waves. This means that it is not possible to have monochromatic $R$ (or $L$) polarized waves traveling in all directions simultaneously; there must be at least one dark spot. Notice that this obstruction to a global circularly polarized basis is a result of the topological nontriviality of the well-defined vector bundles $\zeta_\pm$. In contrast, the obstruction to a global linearly polarized basis was due to the nonexistence of linearly polarized subbundles. 

The geometric separation into $R$ and $L$ waves can be carried out more systematically by using Wigner's little group method \cite{Wigner1939,Weinberg1995}, and this shows how these waves are characterized by their helicity. This method was originally used to characterize the UIRs of the \Poincare group, resulting in Wigner's classification of particles. We can apply the same idea here to classify the quasiparticles in the cold plasma equations. The vector bundle version of the little group method was developed in detail in Refs. \cite{Simms1968,PalmerducaQin_PT}. Here we will present a simplified version of this technique which suffices because the symmetry group $G = \Real^{3+1} \rtimes \SO(3)$ does not contain boosts. Heuristically, the idea is to find an action of a group $H$ which is smaller than $G$, but which still contains enough information to classify the action of $G$.

Consider $\pi:\zeta \rightarrow \pspace$ where $\zeta = \zeta_0$ or $\zeta_{EM}$. The little group at $\kvec$ is defined as the subgroup of $H_{\kvec}$ of homogeneous transformations which leave $\kvec$ invariant, that is,
\begin{equation}
    H_{\kvec} \doteq \{ R \in \SO(3) | R\kvec = \kvec \}.
\end{equation}
It is easy to see that $H_{\kvec}$ consists of all rotations $R_{\kvec}(\varphi)$ about $\kvec$ for $\varphi \in [0,2\pi)$, and thus $H_{\kvec}$ is isomorphic to $\SO(2)$ via $R_{\kvec}(\varphi) \mapsto \varphi$. Since $H_{\kvec}$ does not change $\kvec$, the fiber $\zeta(\kvec)$ is a unitary representation of $\SO(2)$. The generator of this action is $\chi = \phat\cdot \mbf{J}= \khat \cdot \mbf{J}$, which is called the helicity operator. The representations of $\SO(2)$ are particularly simple; they are all 1D and are spanned by the eigenvectors of the generator $\khat\cdot \mbf{J}$. Denote the eigenvectors by $(\kvec,\mbf{E}_n)$ and eigenvalues as $h_n$ so that
\begin{equation}
    (\khat \cdot \mbf{J})(\kvec,\mbf{E}_n) = h_n(\kvec,\mbf{E}_n).
\end{equation}
Exponentiating gives
\begin{equation}
    \Sigma\big(R_{\kvec}(\varphi)\big)(\kvec,\mbf{E}) = e^{-ih_n\varphi}(\kvec,\mbf{E}).
\end{equation}
The condition that $\Sigma\big(R_{\kvec}(2\pi)\big) = 1$ implies that $h_n$ must be an integer. Since the integers are discrete, they cannot vary with $\kvec$. Thus, the vector space spanned by $(\kvec,\mbf{E}_n)$ smoothly varies with $\kvec$ and these vector spaces fit together to form a line bundle $\zeta_{h_n}$ (assuming no two $h_n$ are degenerate). Thus, we can decompose $\zeta$ as
\begin{equation}
    \zeta = \zeta_{h_1} \oplus \cdots \oplus \zeta_{h_r}.
\end{equation}
Furthermore, it is straightforward to check explicitly that the helicity operator $\khat \cdot \mbf{J}$ commutes with all the generators of $G$, and thus the $\zeta_{h_n}$ are irreducible bundle representations of $G$. Applying this to the case of $\zeta_{EM}$, we note that if $(\uv{e}_1,\uv{e}_2,\khat)$ is a right-handed orthonormal basis of $\Real^3$, then
\begin{equation}
    \Sigma\big(R_{\kvec}(\varphi)\big)(\kvec,\uv{e}_1\pm i\uv{e}_2) = e^{\mp i\varphi}(\kvec,\uv{e}_1\pm i\uv{e}_2),
\end{equation}
so the helicities are $\pm1$. This shows that the $R$ and $L$ bundles, $\zeta_\pm$, defined earlier result from the little group method as the helicity $\pm1$ bundles. As the notation $\zeta_0$ for the ES bundle suggests, it is the helicity zero bundle since the modes $(\kvec,E\khat)$ are invariant under the action of $R_{\kvec}(\varphi)$. The upshot is that the little group method furnishes the helicity operator $\mbf{J}\cdot\khat$ which lifts the internal two-fold degeneracy of the EM waves.

\subsection{The electromagnetic-electrostatic resonance as a precondition for topological nontriviality}
The electromagnetic dispersion relation $\omega_\pm = \sqrt{\omega_p^2 + k^2c^2}$ is the same as that of a relativistic particle with mass $m_0 = \hbar \omega_p/c^2$. This can be understood as the inertia of the plasma giving the electromagnetic waves an effective mass. Massive particles are typically associated with trivial topology, and for this reason it is surprising that we find $\zeta_\pm$ are topologically nontrivial. Indeed, for elementary massive particles the momentum space is the non-punctured mass hyperboloid \cite{PalmerducaQin_PT}
\begin{equation}
    \bar{\mathcal{M}}_{m_0} = \Big\{(\omega/c,\kvec) \in \Real^{3+1}\Big| \omega = \sqrt{|\kvec|^2c^2 + \frac{m^2c^4}{\hbar^2}} \Big\}.
\end{equation}
This is a contractible manifold, meaning that it can be smoothly deformation retracted onto a single point. As every vector bundle over a contractible base manifold is topologically trivial (\cite{Bott1982}, Corollary 6.9), the vector bundles describing the states of elementary massive particles are trivial. Massless particles, such as vacuum photons, are not subject to this constraint since their momentum space is given by the forward lightcone
\begin{equation}
    \Lightcone = \Big\{(\omega/c,\kvec) \in \Real^{3+1}\setminus \{(0,\mbf{0})\} | \omega = |\kvec|c \Big\}
\end{equation}
with the vertex at $\omega = 0$, $\kvec = 0$ removed since massless particles have no rest frame. Note that $\Lightcone \cong \pspace$, since it can be parameterized by just $\kvec$. $\Lightcone$ is not contractible due to this hole at the origin, and as such this momentum space can support topologically nontrivial particle bundles such as the $R$ and $L$ photon bundles $\gamma_\pm$.

The reason the electromagnetic plasma bundles $\zeta_\pm$ can be topologically nontrivial is due to the resonance with the electrostatic bundle $\zeta_0$ at $(\omega/c,\kvec) = (\omega_p/c,\mbf{0})$. The electrostatic modes can be thought of as longitudinal photons in plasma \cite{Qin2024plasma}. In the vacuum, such longitudinal modes are nonphysical as they violate Gauss's law, but these modes reappear once an effective mass is introduced. As discussed in Sec. \ref{sec:plasma:Vector_bundle_structure}, the point of resonance must be removed from the base manifolds of $\zeta_\pm$ and $\zeta_0$ to form well-defined vector bundles. This creates a hole in momentum space which allows for the possibility of topologically nontrivial behavior of the plasma waves. 

\section{Angular momentum eigenmodes, spin-weighted spherical harmonics, and the nonexistence of SAM and OAM operators}\label{sec:plasma:AM_of_plasma_waves}
We now consider the following question: what are the eigenmodes of the angular momentum operator for plasma waves? The angular momentum operator $\mbf{J}$ is the generator of the rotational symmetry, so we will restrict our focus to the symmetry group $\SO(3)$. It is helpful to first separate the radial and angular variations in momentum space since only the latter is important for studying the angular momentum: (\cite{Reed1975}, p. 160)
\begin{equation}
    L^2(\zeta_h) = L^2(\Real^+) \otimes L^2(\zeta_h|_{S^2}).
\end{equation}
Here, $L^2(\Real^+)$ is the space of square-integrable functions of the radial coordinate $|\kvec|$ with respect to the measure $|\kvec|^2d|\kvec|$ and $\zeta_h|_{S^2}$ is the bundle of monochromatic waves obtained from $\zeta$ by restricting the base manifold $\pspace$ to the unit sphere $S^2$ in momentum space. This expresses the fact that any field configuration $\mbf{E}_h(\kvec) \in \zeta_h$ can be written as a sum of terms of the form $E(|\kvec|){\mbf{\alpha}}_h(\khat)$ where $\mbf{\alpha}_h \in L^2(\zeta_h|_{S^2})$. The angular momentum operator $\mbf{J}$ only acts nontrivially on $L^2(\zeta|_{S^2})$ since $|\kvec|$ is invariant under rotations. 

The Hilbert space $\Gamma(\zeta_h|_{S^2})$ is an infinite-dimensional unitary representation of $\SO(3)$ for $h = 0,\pm1$. Since the UIRs of $\SO(3)$ are all finite-dimensional, as is true of any compact group (cf. \cite{Johnson1976}, Theorem 3.9), $\Gamma(\zeta_{h}|_{S^2})$ must decompose into an infinite number of UIRs. The (non-projective) vector space UIRs of $\SO(3)$ are classified by the nonnegative integers $n$ and have dimension $2n+1$; this follows from the usual ladder operator argument \cite{Hall2013}. This means that there must exist multiplets of monochromatic field configurations $\mbf{E}_{hjm}(\khat)$
which are closed under rotations, where $h$ and $j$ label a multiplet and $m$ spans all integers between $-j$ and $j$. That is, if $R$ is a rotation, then $\tilde{\Sigma}_R\mbf{E}_{hjm}$ is a linear combination of the $\mbf{E}_{hjm'}$ for $|m'|\leq j$. Indeed, by the representation theory of $\SO(3)$, these multiplets solve the equations
\begin{subequations}
\label{eq:plasma:AM_bundle}
\begin{align}
    J_3 \mbf{E}_{hjm} &= m\mbf{E}_{hjm}, \\
    J_\pm \mbf{E}_{hjm} &= [(j \mp m)(j + 1 \pm m)]^{1/2} \,  \mbf{E}_{hjm}, \\
    J^2 \mbf{E}_{hjm} &= j(j+1) \mbf{E}_{hjm}
\end{align}
\end{subequations}
where $J_\pm = J_1 \pm iJ_2$ and $J^2 = J_1^2 + J^2_2 + J^2_3$. Note that if $\mbf{E}_{hjm}$ were a scalar function rather than a section of a vector bundle, then these equations would be equivalent to the angular part of the time-independent Schrödinger equation, which is solved by the ordinary spherical harmonics. In fact, this is precisely the solution for Langmuir waves since such waves are effectively scalar. Indeed, if we fix $h=0$, then we can write $\mbf{E}_{0jm}(\khat) = E_{0jm}(\khat)\khat$. Since the vector field $\khat$ is invariant under all rotations, it follows that $J_a \khat = 0$ for any $a$. Thus,
\begin{subequations}\begin{align}
    (J_a\mbf{E}_{0jm})(\khat) &= \frac{d}{d\varphi}\Big|_{\varphi=0} R_a(\varphi)\mbf{E}_{0jm}\big(R_a(-\varphi)\khat\big) \\
    &= \khat \frac{d}{d\varphi}\Big|_{\varphi=0} E_{0jm}\big(R_a(-\varphi)\khat\big) \\
    &=  (L_aE_{0jm})\khat
\end{align}\end{subequations}
where
\begin{equation}
\mbf{L} = -i\Big(\ephi \partial_\theta - \etheta \frac{1}{\sin\theta} \partial_\phi\Big)
\end{equation}
is the angular momentum operator on functions (\cite{Zettili2009}, Eq. B.24) and $L_a  = \uv{e}_a \cdot \mbf{L}$. For $h=0$, Eq. (\ref{eq:plasma:AM_bundle}) then reduces to
\begin{subequations}
\label{eq:plasma:AM_ES}
\begin{align}
    L_3 E_{0jm} &= mE_{0jm}, \\
    L_\pm E_{0jm} &= [(j \mp m)(j + 1 \pm m)]^{1/2} \,  E_{0jm}, \\
    L^2 E_{0jm} &= j(j+1) E_{0jm}.
\end{align}
\end{subequations}
These are solved by the ordinary spherical harmonics
\begin{equation}\label{eq:plasma:electrostatic_SH}
    E_{0jm}(\khat) = Y_{jm}(\theta,\phi).
\end{equation}
Langmuir waves can thus be expanded in terms of the spherical harmonics, and the labels $j$ and $m$ describe the total and $z$ angular momentum, respectively. 

The situation is more interesting for EM waves with $h= \pm1$ due to the nontrivial vector polarizations. We will work in a particular basis for $R$ and $L$ waves, namely,
\begin{equation}
    \eh = \frac{1}{\sqrt{2}}(\etheta +  i\,\sign(h)\ephi).
\end{equation}
This basis will allow us to express the angular momentum basis in terms of a familiar set of special functions. Note that this basis has discontinuities at the poles, thus, if we write
\begin{equation} \label{eq:plasma:hjm_expansion}
    \mbf{E}_{hjm}(\khat) = E_{hjm}(\khat)\eh
\end{equation}
where $\mbf{E}_{hjm}(\khat)$ is everywhere smooth, then $E_{hjm}(\khat)$ will be discontinuous at the poles (unless $\mbf{E}_{hjm}$ vanishes at the poles); importantly, these are coordinate singularities in the representation, not true singularities of $\mbf{E}_{hjm}$. We also define
\begin{equation}
    \uv{e}_0 = \khat
\end{equation}
so that the following analysis also applies to the $h=0$ case. In this basis, we can write
\begin{subequations}\begin{align}
    J_3 \mbf{E}_{hjm} &= (J_{h3}'E_{hjm}) \eh \\
    J_\pm \mbf{E}_{hjm} &= (J_{h\pm}'E_{hjm}) \eh \\
    J^2 \mbf{E}_{hjm} &= (J_h'^2E_{hjm}) \eh
\end{align}\end{subequations}
for some operators $\mbf{J}'_h$ on scalar functions. The same set of equations (for arbitrary integer $h$) appeared in our analysis of the angular momentum eigenstates of massless particles \cite{PalmerducaQin_SWSH}, where it was shown that
\begin{subequations}\label{eq:plasma:multiplet_components}
\begin{align}
    J_{hz}' &= -i \partial_\phi, \\
    J_{h\pm}' &= e^{\pm i \phi}\Big(\pm \partial_\theta + i\frac{\cos \theta}{\sin \theta} \partial_\phi + \frac{h}{\sin \theta} \Big), \\
    J'^2_{h} &= -\nabla^2 - \frac{2 h\cos \theta}{\sin^2 \theta}L_z + \frac{h^2}{\sin ^2 \theta}.
\end{align}
\end{subequations}
The $E_{hjm}$ thus solve
\begin{subequations}
\begin{align}
    J_{hz}'  E_{hjm} &= m E_{hjm}, \\
    J_{h\pm}'  E_{hjm}&= [(j \mp m)(j + 1 \pm m)]^{1/2}   E_{hjm}, \\
    J'^2_{h}  E_{hjm} &= j(j+1) E_{hjm}.
\end{align}
\end{subequations}
Dray \cite{Dray1985} showed that these equations are solved by the SWSHs $_{-h}{Y}_{jm}$ rather than the ordinary spherical harmonics. The SWSHs are given explicitly by \cite{Goldberg1967}
\begin{align}
    &_h{Y}_{jm}(\theta,\phi) = \Big[ \frac{(j + m)! (j-m)! (2j+1)}{4\pi (j+h)!(j-h)!}\Big]^{\frac{1}{2}} (\sin \theta / 2)^{2j} \nonumber\\ 
    &\times \sum_{q}\binom{j-h}{q}\binom{j+h}{q+h-m}(-1)^{j-q-h-m}e^{im\phi}(\cot \theta / 2)^{2q+h-m}
\end{align}
where $j \geq |h|$, $-j \leq m \leq j$ and the sum ranges from $q = \max (0,m-h)$ to $q = \min (j-h,j+m)$. These are special functions which were first defined by Newman and Penrose in their study of asymptotically flat spacetimes \cite{Newman1966}, but have since found applications in many other areas including the interaction of charged particles with magnetic monopoles \cite{Wu1975,Wu1976,Dray1985,Fakhri2007}, the study of complex spacetimes \cite{Curtis1978}, analysis of the cosmic microwave background radiation \cite{Wiaux2006}, in geophysical \cite{Michel2020} and computer graphics applications \cite{Yi2024}, and in the analysis of the angular momentum of massless particles \cite{PalmerducaQin_SWSH}. This appears to be the first time the SWSHs have shown up in the context of plasma physics.

The simultaneous eigenstates of $J^2$ and $J_z$ in $L^2(\zeta_h|_{S^2})$ are thus given by
\begin{equation}
    \mbf{E}_{hjm}(
\khat) = {_{-h}{Y}_{jm}}(\theta,\phi)\eh
\end{equation}
Note that when $h = 0$, the SWSHs reduce to the ordinary spherical harmonics, $_0{Y}_{jm} = Y_{jm}$, so we recover the ES result in Eq. (\ref{eq:plasma:electrostatic_SH}). For fixed $h$, the SWSHs are orthogonal \cite{Dray1985},
\begin{equation}
    \int_{S^2} \,_h{Y}^*_{jm}\,_h{Y}_{j'm'} \sin\theta \,d\theta \,d\phi =\delta_{jj'} \delta_{mm'},
\end{equation}
so that any $\mbf{E}(\kvec)\in L^2(\zeta_+) \oplus L^2(\zeta_0) \oplus L^2(\zeta_-)$ is easily expanded in terms of SWSHs:
\begin{gather}
    \mbf{E}(\kvec) = \sum_{h = -1}^1\sum_{j=|h|}^\infty \sum_{|m|\leq j}\alpha_{hjm}(|\kvec|) {_{-h}Y_{jm}}(\theta,\phi)\eh, \\
    \alpha_{hjm}(|\kvec|) = \int {_{-h}Y^*_{jm}} (\theta,\phi) \eh^* \cdot\mbf{E}(\kvec) \sin\theta \, d\theta \, d\phi.
\end{gather}
Note that this gives a countable basis for monochromatic waves, as an arbitrary monochromatic wave $\mbf{E}(\khat)$ is specified by a countable set of complex numbers $\alpha_{hjm}$. Compare this, for example, to the expansion $\mbf{E}(\khat) = \sum_hE_h(\khat)\eh$ in terms of the momentum eigenstates $\eh$---in this case the state is specified by an uncountable set of numbers $E_h(\khat)$ as $\khat$ spans over the unit sphere. We have essentially followed the technique frequently employed in quantum mechanics of lifting degeneracies using commuting operators. In this case, the $_h Y_{jm}\eh\delta(|\kvec|-|\kvec_0|)$ are simultaneous eigenvectors of the energy $H$, helicity $\chi$, $J^2$, and $J_3$. Indeed, since we can describe any wave using these eigenstates, these four operators give a complete set of commuting operators (CSCO) for plasma waves, that is, a basis for the states is determined by the eigenvalues $(\omega_h(|\kvec|),h,j,m)$. Compare this to a massive particle with spin and orbital degrees of freedom, for which a CSCO is given by the \emph{five} operators $H,J^2,J_3,S^2,L^2$ \cite{Shankar_QM}. $S^2$ and $h$ are determined by the identity of the particle (\emph{i.e.}, by the UIR of the symmetry group), but in the massive case it is necessary to specify both $J^2$ and $L^2$ (assuming nonzero spin), while in the massless case $J^2$ suffices. This observation suggests that there may be issues with defining SAM and OAM operators for EM plasma waves, a topic that we will examine in greater detail.

The form of the SWSHs reveals some interesting features of the rotational symmetry and angular momentum of plasma waves. We note first that there are no SWSHs with $j<|h|$, showing that $|h|$ is a lower bound on the total angular momentum, in particular,
\begin{equation}
    \frac{\langle\mbf{E}_h|J^2|\mbf{E}_h\rangle}{|\mbf{E}_h|^2} \geq |h|(|h|+1).
\end{equation}
Thus, $R$ and $L$ waves have no $j = 0$ state, reflecting the fact that there is no wave $\mbf{E}_\pm(\khat)$ that is rotationally invariant. This agrees with our topological analysis of $R$ and $L$ waves, for such a rotationally invariant $\mbf{E}_\pm(\khat)$ would have to be nonvanishing for all $\khat$, which is prohibited by the topological nontriviality of $\zeta_\pm$. Notice that this lack of $j = 0 $ states is not consistent with the existence of a standard SAM-OAM decomposition of the angular momentum. For example, consider a massive spin $1$ particle, with the $\SO(3)$ symmetry of its spin space described by the representation $V_1$ and its orbital space by 
\begin{equation}\label{eq:plasma:spin_1_orbital_decomp}
    \mathcal{H}_o = V_0 \oplus V_1 \oplus V_2 \oplus\cdots.
\end{equation}
Here, $V_l$ denotes the canonical $\SO(3)$ representations with dimension $2l+1$, and in Eq. (\ref{eq:plasma:spin_1_orbital_decomp}) they are spanned by the ordinary spherical harmonic multiplets $Y_{lm}$ with $|m|\leq l$. The total angular momentum multiplet structure is determined by the addition of angular momentum formula, where for $a\geq b$,
\begin{equation}\label{eq:plasma:AM_addition}
    V_a \otimes V_b \cong V_{a-b} \oplus V_{a-b + 1} \oplus \cdots \oplus V_{a+b}.
\end{equation}
The complete Hilbert space for a spin $1$ particle is then
\begin{equation}\label{eq:plasma:spin_1_decomp}
    V_1 \otimes (V_0 \oplus V_1 \oplus \cdots) = V_0 \oplus 3V_1 \oplus 3V_2 \oplus \cdots.
\end{equation}
In particular, the copy of $V_0$ shows that there is a total angular momentum $0$ state. In general, if the total orbital angular momentum can assume the same value $s$ as the total spin angular momentum, as is true for all massive particles particles, then there must exist a total angular momentum $0$ multiplet resulting from $V_{s} \otimes V_{s}$. $R$ and $L$ plasma waves possess no such multiplet, further suggesting that plasma waves do not admit an SAM-OAM splitting. Additional support for this conclusion comes from analyzing the full multiplet structure of $R$ and $L$ plasma waves. Indeed, we see that for $h = \pm1$,
\begin{equation}\label{eq:plasma:plasma_SO3_decomp}
    \Gamma(\zeta_{\pm}|_{S^2}) = V_1 \oplus V_2 \oplus V_3 \oplus \cdots
\end{equation}
where each $V_j$ corresponds to the multiplet $_{\mp1}Y_{jm}\epm$. There is only one multiplet for each $j \geq 1$. This is inconsistent with the typical multiplet structure of particles with an SAM-OAM decomposition, which typically produce higher multiplicities as in Eq. (\ref{eq:plasma:spin_1_decomp}). This argument is highly suggestive, but not fully rigorous since it is possible to contrive exotic OAM multiplet structures with missing OAM states which are consistent with Eq. (\ref{eq:plasma:plasma_SO3_decomp}) \cite{PalmerducaQin_SWSH}, for example,
\begin{equation}
    V_1 \otimes (V_0 \otimes V_3 \otimes V_6 \otimes \cdots) = V_1 \oplus V_2 \oplus V_3 \oplus \cdots.
\end{equation}
However, such an OAM multiplet structure is very unnatural, with the OAM label $l$ restricted to integer multiples of $3$, a situation that does not show up in any standard physics applications.

We can rigorously show that no SAM-OAM decomposition is possible, ruling out such exotic OAM structures, by essentially repeating an argument that applies to massless particles (cf. \cite{PalmerducaQin_SAMOAM}, No-Go Theorem 1). Assume $\mbf{J} = \mbf{S} + \mbf{L}$ where $\mbf{S}$ and $\mbf{L}$ commute with the Hamiltonian and generate $\SO(3)$ symmetries. Assume further that $\mbf{S}$ generates an internal symmetry, meaning that it does not alter $\kvec$. Since they commute with the Hamiltonian, $\mbf{S}$ and $\mbf{L}$ do not mix $L^2(\zeta_{EM})$ and $L^2(\zeta_{0})$, and thus can be considered to act on EM and Langmuir waves separately. Since $\mbf{S}$ is internal, it is an operator on each fiber $\zeta_{0}(\kvec)$ and $\zeta_{EM}(\kvec)$. Since $\mbf{S}$ generates $\SO(3)$, the fibers $\zeta_{0}(\kvec)$ and $\zeta_{EM}(\kvec)$ are vector space representations of $\SO(3)$ of dimension $1$ and $2$, respectively. We can decompose these into UIRs of $\SO(3)$, which are isomorphic to the complex vector spaces $V_n$ with dimension $2n+1$. $\zeta_0$ is one-dimensional, so it must be equivalent to the trivial representation $V_0$ whose generator is $0$. Similarly, there are no non-projective representations of dimension $2$, so $\zeta_{EM}(\kvec) \cong V_0 \oplus V_0$ and therefore also corresponds to a trivial representation with generator $0$. Thus, we find that $\mbf{S} = 0$ and therefore there is no nontrivial SAM-OAM decomposition.

\section{Helicity and orbital quasi-angular momenta}\label{sec:plasma:plasma_par_perp_decomp}
We have shown in the previous section that there is no true SAM-OAM decomposition for cold plasma waves. By this we mean that if one splits $\mbf{J}$ into two operators $\mbf{S}$ and $\mbf{L}$, then these operators do not generate 3D rotations, or equivalently, they will not satisfy the angular momentum commutation relations. This has significant implications, because the majority of the theoretical tools pertaining to SAM and OAM rely on their $\SO(3)$ structure. For example, the addition of angular momentum formula, the Clebsch-Gordan basis, and the Casimir invariance of $S^2$  and $L^2$ do not hold in the absence of $\SO(3)$ symmetry. Furthermore, we are accustomed to $S_z$, $S^2$, $L_z$, and $L^2$ having discrete spectra; this too may fail if we consider $\mbf{S}$ and $\mbf{L}$ which do not generate rotational symmetries.

However, we now show that, as in the case of photons \cite{Bliokh2010,PalmerducaQin_PT,PalmerducaQin_connection}, it is possible to decompose $\mbf{J}$ into two well-defined vector operators defined purely in terms of symmetry generators:
\begin{subequations}\label{eq:plasma:par_perp_splitting}
\begin{align}
    \mbf{J} &= \mbf{J}_\parallel + \mbf{J}_{\perp} \\
    \mbf{J}_\parallel &=  (\mbf{J}\cdot \phat) \phat  = \chi\phat\\
    \mbf{J}_\perp &= -\phat \times (\phat \times \mbf{J})
\end{align}
\end{subequations}
where $\phat = \mbf{P}/|\mbf{P}|$. That they are vector operators means that they satisfy \cite{Hall2013}
\begin{align}
    [J_{||,a}, J_{b}] &= i\epsilon_{abc}J_{\parallel,c} \\ 
    [J_{\perp,a}, J_{b}] &= i\epsilon_{abc}J_{\perp,c}.
\end{align}
We will call $\mbf{J}_\parallel$ and $\mbf{J}_\perp $ the helicity quasi-angular momentum and orbital quasi-angular momentum, respectively. The term quasi-angular momentum reflects the fact that these operators cannot be generators of rotation. While the analogous splitting has been previously applied to photons in vacuum \cite{VanEnk1994_EPL_1,Bliokh2010,PalmerducaQin_PT,PalmerducaQin_connection}, this is the first time this operator splitting has been given for electromagnetic waves in plasmas. The validity of this splitting can be established as follows. If $\mbf{A}$, $\mbf{B}$, $\mbf{C}$ are vector operators such that $\mbf{A}$ and $\mbf{B}$ commute, then \cite{PalmerducaQin_connection}
\begin{equation}
    \mbf{A} \times (\mbf{B}\times\mbf{C})  = \mbf{B}(\mbf{A}\cdot\mbf{C}) - (\mbf{A}\cdot\mbf{B})\mbf{C}.
\end{equation}
Setting $\mbf{A}=\mbf{B}=\phat$ and $\mbf{C} = \mbf{J}$ and using the fact that $(\mbf{J}\cdot\phat)$ commutes with $\phat$ gives Eq. 
(\ref{eq:plasma:par_perp_splitting}). 

While it is not difficult to check that this splitting is technically valid, it can in fact be derived from a more fundamental theoretical consideration---rotational symmetry. We recently showed, in the context of elementary particles, that this splitting arises from considering connections on vector bundles \cite{PalmerducaQin_connection, Tu2017differential}, the same type of mathematical objects which are ubiquitous in general relativity. We showed that every connection $\mbf{D}$ (which can be thought of as a covariant derivative) on a vector bundle over momentum space induces a splitting of the angular momentum of the form
\begin{subequations}\label{eq:plasma:cx_splitting}
\begin{align}
    \mbf{J} &= \mbf{L}^{\mbf{D}} + \mbf{S}^{\mbf{D}} \\
    \mbf{L}^{\mbf{D}} &\doteq -i\mbf{P} \times\mbf{D} \\
    \mbf{S}^{\mbf{D}} &\doteq \mbf{J} - \mbf{L}^{\mbf{D}}
\end{align}
\end{subequations}
Notice that this is what one would expect from the semiclassical heuristic of taking $\mbf{L} = \mbf{x}\times\mbf{P}$ and making the replacement $\mbf{x}\rightarrow i\mbf{D}$. However, $\mbf{L}^{\mbf{D}}$ and $\mbf{S}^{\mbf{D}}$ satisfy $\SO(3)$ commutation relations if and only if $\mbf{D}$ has vanishing curvature (\cite{PalmerducaQin_connection}, Theorem 3), which is not possible for the EM bundles $\zeta_\pm$ since they are topologically nontrivial. Nevertheless, there is a unique connection $\mbf{D}$ on $\zeta_{\pm}|_{S^2}$ induced by their rotational symmetry, namely,
\begin{equation}
    \boldsymbol{D} = -i\frac{\hat{\mbf{P}}\times \mbf{J}}{|\mbf{P}|},
\end{equation}
and this induces the splitting into the orbital and helicity quasi-angular momenta:
\begin{gather}
    \mbf{L}^{\mbf{D}} = -\phat \times(\phat \times\mbf{J}) = \mbf{J}_\perp \\
    \mbf{S}^{\mbf{D}} = \mbf{J} - \mbf{L}^{\mbf{D}} = \mbf{J}_\parallel.
\end{gather}
Note that the Langmuir bundle is topologically trivial and $\boldsymbol{D}$ is flat on this bundle. However, the induced splitting is trivial in this case since it has helicity $\chi = 0$ and thus $\mbf{J}_\parallel = 0$ and $\mbf{J}_\perp = \mbf{J}$. We have thus shown that the quasi-angular momentum decomposition arises from the same general theory which describes the corresponding decomposition for massless elementary particles as well as the standard SAM-OAM splitting for massive particles \cite{PalmerducaQin_connection}.

The quasi-angular momentum operators have a number of features which distinguish them from true angular momentum operators. Most importantly, they do not satisfy angular momentum commutation relations, but rather the peculiar relations
\begin{subequations}\label{eq:plasma:par_perp_comm_relations}
\begin{gather}
    [J_{\parallel,a},J_{\parallel,b}] = 0 \neq i\epsilon_{abc}J_{\parallel,c} \\
    [J_{\perp,a}, J_{\perp,b}] = i\epsilon_{abc}(J_{\perp,c} - J_{\parallel,c})\neq i\epsilon_{abc}J_{\perp,c}.\label{eq:plasma:J_perp_comm}
\end{gather}
\end{subequations}
The commutation relations for the components of $J_{\parallel}$ show that they actually generate a 3D translational $\mathbb{R}^3$ symmetry rather than a 3D rotation symmetry. That the $\mbf{J}_\perp$ components are not closed under the commutator (due to the $\mbf{J}_{\parallel,c}$ term) shows that the vector operator $\mbf{J}_\perp$ does not generate any symmetry at all.

Spectral analysis reveals another strange feature of these operators. In terms of spherical coordinates $(|\kvec|,\theta,\phi)$, 
\begin{subequations}\begin{align}
    J_{\parallel,z}[\delta(\theta-\theta_0)e^{im\phi}\epm] &= \frac{P_z}{|\mbf{P}|}\chi[\delta(\theta-\theta_0)e^{im\phi}\epm]\\
    &=\pm \frac{k_z}{|\kvec|}\delta(\theta-\theta_0)e^{im\phi}\epm \\
    &= \pm \cos(\theta_0)\delta(\theta-\theta_0)e^{im\phi}\epm
\end{align}\end{subequations}
for any integer $m$, showing that
\begin{equation}
    \mbf{\alpha}^\pm_{\theta_0}(\kvec) = \delta(\theta-\theta_0)e^{im\phi}\epm
\end{equation}
is a (generalized \cite{Hall2013}) eigenvector of $J_{\parallel,z}$ with eigenvalue $\pm\cos(\theta_0)$ on $L^2(\zeta_\pm)$. From the fact that $\chi$ and $\hat{\mbf{P}}$ are both bounded operators with norms $||\chi|| = ||\hat{\mbf{P}}|| = 1$, we see that  $\mbf{J}_\parallel$ does not have any eigenvalues of magnitude greater than $1$. $J_{\parallel,z}$ thus has a continuous spectrum given by the interval $[-1,1]$. This is in stark contrast to the components of a true angular momentum operator, which always have discrete integer or half-integer eigenvalues. Furthermore, $J_z\delta(\theta - \theta_0)\mbf{e}_\pm = 0$ since $\delta(\theta - \theta_0)\mbf{e}_\pm$ is invariant about the $k_z$ axis, so it follows that
\begin{align}
    J_z \mbf{\alpha}^\pm_{\theta_0} &= \delta(\theta - \theta_0)\mbf{e}_\pm J_ze^{im\phi}  \\
    &=m\mbf{\alpha}^\pm_{\theta_0},
\end{align}
showing that $\mbf{\alpha}^\pm_{\theta_0}$ is also an eigenvector of $J_z$ with eigenvalue $m$. Thus, the spectrum of $J_{\perp,z} = J_z - J_{\parallel,z}$ contains the interval $[m-1,m+1]$ for each $m$. Since $m$ can be any integer, the spectrum of $J_{\perp,z}$ is all of $\Real$. This again contrasts with the discrete spectral behavior that occurs for true OAM operators.

Nevertheless, operators satisfying the non-$\SO(3)$ commutation relations in Eq. (\ref{eq:plasma:par_perp_comm_relations}) have long been used in the optics literature and are often called SAM and OAM operators \cite{VanEnk1994_EPL_1,Bliokh2010,Leader2014,Leader2016,Leader2019}. They are gauge-invariant vector operators and appear to be important in experimental optics, despite not being true SAM and OAM operators \cite{Leader2016}. We show in the Appendix that the expectation values of $\mbf{J}_\parallel$ and $\mbf{J}_\perp$, as well as those of $H, \mbf{P},$ and $\mbf{J}$, can be calculated by integrals over position or momentum space:
\begin{align}
    \frac{1}{8\pi}\langle \mbf{E} | H |\mbf{E}\rangle  &= \sum_{h}\frac{1}{8\pi}\int \frac{d^3k}{\omega_h}\mbf{E}_h^*(\mbf{k}) \cdot [H\mbf{E}_h](\mbf{k}) \nonumber \\
    &= \sum_{h}\frac{1}{8\pi}\int d^3k\,|\mbf{E}_h(\mbf{k})|^2 \nonumber \\
    &= \int d^3x \, \Big[\frac{1}{8\pi}\big(\mathcal{E}^2(\mbf{x}) + \mathcal{B}^2(\mbf{x})\big) + \frac{1}{2}mn_0v^2(\mbf{x})\Big] \label{eq:plasma:E_expectation}\\
    \frac{1}{8\pi}\langle \mbf{E} | \mbf{P}|\mbf{E}\rangle
    &= \sum_h\frac{1}{8\pi} \int d^3k\frac{\kvec}{\omega_h}|\mbf{E}_h(\mbf{k})|^2 \nonumber\\
    &= \int d^3x \, \Big[\frac{1}{4\pi c}\boldcal{E}(\mbf{x}) \times \boldcal{B}(\mbf{x}) + mn_{1,h}(\mbf{x})\mbf{v}(\mbf{x})\Big] \label{eq:plasma:P_expectation}
\end{align}
\begin{align}
    \frac{1}{8\pi}\langle \mbf{E} | \mbf{J}|\mbf{E}\rangle &= -i\sum_{h}\frac{1}{8\pi}\int \frac{d^3k}{\omega_h}\Big(\sum_{n}E^*_{h,n}\cdot (\kvec \times \nabla_{\kvec})E_{h,n} + \mbf{E}^*_h \times \mbf{E}_h\Big) \nonumber\\
    &= \int d^3x \, \Big[\frac{1}{4\pi c}\mbf{x} \times \big(\boldcal{E}(\mbf{x}) \times \boldcal{B}(\mbf{x})\big) + mn_1(\mbf{x})\mbf{x} \times \mbf{v}\Big] \label{eq:plasma:J_expectation} \\
    \frac{1}{8\pi}\langle \mbf{E} | \mbf{J}_\parallel|\mbf{E}\rangle &= -i\sum_{h}\frac{1}{8\pi}\int \frac{d^3k}{\omega_h}\mbf{E}^*_h \times \mbf{E}_h\nonumber\\
    &= \frac{1}{4\pi c}\int d^3x \, \boldcal{E}^{\perp}(\mbf{x}) \times \boldcal{A}^{\perp}(\mbf{x}) \label{eq:plasma:J_par_expectation} \\
    \frac{1}{8\pi}\langle \mbf{E} | \mbf{J}_\perp|\mbf{E}\rangle &= -i\sum_{h,n}\frac{1}{8\pi}\int \frac{d^3k}{\omega_h}E^*_{h,n} (\kvec \times \nabla_{\kvec})E_{h,n}\nonumber\\
    &= \int d^3x \Big[ \, \frac{1}{4\pi c}\sum_n \mathcal{E}^{\perp}_{n} (\mbf{x}\times \nabla) \mathcal{A}^\perp_{n} + mn_{1}\mbf{x} \times \mbf{v}\nonumber\\
    & \qquad\qquad\qquad -\frac{e}{c}n_{1}\mbf{x} \times \boldcal{A}^\perp \Big]\label{eq:plasma:J_perp_expectation}
\end{align}
where $\omega_0 = \omega_p$, $\omega_\pm = \sqrt{\omega^2_p + |\kvec|^2}$, and $\mbf{E}_h = \eh^*\cdot\mbf{E}$. $\boldcal{A}^\perp$ is the transverse vector potential, the divergence-free part of the Helmholtz decomposition of the vector potential. It is the component of the vector potential generating the EM waves $\mbf{E}_\pm$, but not the Langmuir waves $\mbf{E}_0$. The transverse electric field is given by $\boldcal{E}^\perp = c^{-1}\partial_t\boldcal{A}^\perp$. Unlike the full vector potential, $\boldcal{A}^\perp$ is gauge-invariant, so the expectation values of $\mbf{J}_\parallel$ and $\mbf{J}_\perp$ are also gauge-invariant and thus physically measurable \cite{VanEnk1994_EPL_1}. In the Coulomb gauge, we simply have $\boldcal{A} = \boldcal{A}^\perp$. These expectation values of $\mbf{J}_\parallel$ and $\mbf{J}_\perp$ agree with expressions given recently by Bliokh and Bliokh (\cite{Bliokh2022}, cf. Table 1). We note that Ref. \cite{Bliokh2022} referred to these as the spin and orbital components of the canonical angular momentum, but as we have shown, these do not correspond to generators of rotational symmetries, and are thus not true angular momenta. We also note that, unlike the expressions in Ref. \cite{Bliokh2022}, Eqs. (\ref{eq:plasma:E_expectation})-(\ref{eq:plasma:J_perp_expectation}) do not assume that the waves are monochromatic or time-averaged. Bliokh and Bliokh show that $\boldsymbol{J}_\parallel$ is related to the induced plasma magnetization in the inverse Faraday effect, which illustrates that the quasi-angular momentum decomposition is experimentally relevant in plasma physics, despite it not being a true SAM-OAM splitting. This decomposition is also useful for theoretical calculations, and in fact was used in Ref. \cite{PalmerducaQin_connection} to calculate the expressions for $J'_z,$ $J'^2$, and $J'_\pm$ in Eq. (\ref{eq:plasma:multiplet_components}). In summary, the quasi-angular momentum operators are naturally induced by the rotational connection of cold plasma waves, and can be relevant in both experimental and theoretical contexts. It is important, however, to bear in mind that these are not genuine SAM and OAM operators, and thus most of the theoretical tools used for angular momentum operators do not apply to them.

\section{Extension to kinetic plasma models}\label{sec:plasma:kinetic}
We indicate in this section how one might generalize our methods to the treatment of kinetic plasmas. The true (generalized) eigenmodes of a kinetic plasma are the Case-van Kampen modes which have a continuous spectrum \cite{Stix1992,Dodin_WaveNotes}. The continuous spectrum makes these modes difficult to treat using the vector bundle formalism since it would produce infinite-dimensional fibers. Since vector bundles are typically defined to have finite-dimensional fibers, and since many vector bundle methods require this constraint, treating the Case-van Kampen modes would require a modified formalism. While this may be possible, it is not clear that it would be particularly helpful. Case-van Kampen modes are non-normalizable and correspond to nonphysical initial conditions (\cite{Stix1992}, \S8-10; \cite{Dodin_WaveNotes}, \S9.1), and thus are not particularly convenient for treating many physical problems. The standard and often more useful formalism is the Landau initial value approach \cite{Landau1946, Stix1992, Dodin_WaveNotes}, which allows one to focus on plasma configurations corresponding to smooth initial conditions. Following the treatment in Ref. \cite{Dodin_WaveNotes}, by linearizing around a spatially homogeneous distribution function $f_0(\mbf{v})$, one can express the wave equation in terms of the plasma dispersion tensor
\begin{equation}
    \mbf{D}_{E}(\omega,\mbf{k}) = \mbf{N}\mbf{N}^T - \mathds{1}N^2 + \mbf{\epsilon}(\omega,\kvec)
\end{equation}
where $\mbf{\epsilon}$ is the dielectric tensor of the plasma model and $\mbf{N} = c\kvec / \omega$ is the index of refraction tensor. The dispersion relation is given by
\begin{equation}
    \det \mbf{D}_E(\omega,\kvec) = 0.
\end{equation}
Denoting the solution branches by $\omega_n(\kvec)$, the quasimodes $\mbf{E}_n(\kvec)$ are determined by
\begin{equation}
    \det \mbf{D}_E(\omega_n(\kvec),\kvec)\mbf{E}_n(\kvec) = 0.
\end{equation}
The term quasimode is used to distinguish these from the Case-van Kampen modes. The problem thus takes the same form as Eq. (\ref{eq:fluid_dispersion_tensor}) and we see that the quasimodes will globalize to form vector bundles as $\kvec$ is varied. Thus we can study the topology of the quasimodes by the same formalism as we have applied to the cold fluid model. One minor difference is that the $\omega_n$ can generally be complex, reflecting the fact that quasimodes can be damped. This does not cause obvious issues since the base manifold can always be parameterized by $\kvec$.

Consider the example of an unmagnetized isotropic plasma, so that $f_0(\mbf{v}) = f_0(v)$ where $v = |\mbf{v}|$. As in the case of our treatment of the cold fluid plasma, we will ignore the ion contribution to the dispersion tensor. If we pick an orthonormal basis $(\uv{e}_x,\uv{e}_y,\khat)$ at a fixed $\kvec$, then (\cite{Dodin_WaveNotes}, Eq. (10.14))
\begin{equation}
    \mbf{D}_E =
    \begin{pmatrix}
        \epsilon_\perp - N^2 & 0 & 0 \\
        0 & \epsilon_\perp - N^2 & 0 \\
        0 & 0 &  \epsilon_\parallel
    \end{pmatrix}
\end{equation}
where $\mbf{N} = c\kvec /\omega$ and
\begin{gather}
    \epsilon_\parallel(\omega,\kvec) = 1 - \frac{\omega_{p}^2}{k^2}\int_L dv \frac{f_0'(v)}{v - \omega/k}\\
    \epsilon_\perp(\omega,\kvec) = 1 - \frac{\omega^2}{\omega_p^2}\Big( 1 + \int_L d\mbf{v} \frac{k v_y^2}{\omega - kv_x}\frac{\partial f_0}{\partial v_x} \Big).
\end{gather}
The integral is taken over the the Landau contour $L$, which bends under the pole. Thus, there is still a clear distinction between the ES modes $\epsilon_{\parallel}(\omega,\kvec) = 0$, which are longitudinally polarized with $\mbf{E} \times \kvec = 0$, and EM modes $\epsilon_\perp - N^2 = 0$, which are transversely polarized with $\mbf{E} \cdot \kvec = 0$. We also need to consider the set of Weyl points where ES and EM modes degenerate. For concreteness, let us take the background to be Maxwellian with electron thermal velocity $v_T$:
\begin{equation}
    f_0(v) = \frac{1}{\sqrt{2\pi}v_T} e^{-\frac{v^2}{2v_T^2}}.
\end{equation}
Then, if we neglect relativistic corrections $(v_T \ll c)$, the EM dispersion relation is still given by (\cite{Dodin_WaveNotes}, Eq. (10.20))
\begin{equation}
    \omega_{EM}^2 \approx \omega_p^2 + k^2 c^2.
\end{equation}
For small $k$ ($k \ll \omega/v_T$) the real part of the ES frequency is given by (\cite{Dodin_WaveNotes}, Eq. (11.35))
\begin{equation}
    \omega_{ES,r}^2 \approx \omega_p^2 + 3k^2v_T^2.
\end{equation}
Thus, the modes will still only degenerate at $\kvec = 0$. Thus, we see that in these limits, $\zeta_{ES}^K$ and $\zeta_{EM}^K$ are vector bundles over the momentum space $\pspace$ which are topologically equivalent to the cold plasma bundles $\zeta_{ES}$ and $\zeta_{EM}^K$. This suggest that moving to a kinetic model does not necessarily modify the topology. Furthermore, these waves will still be rotationally symmetric. Thus, it appears that the description of the angular momentum eigenmodes in terms of SWSHs, the nonexistence of an SAM-OAM decomposition, and the quasi-angular momentum splitting should still apply in a kinetic unmagnetized plasma. However, we have made a number of assumptions, such as neglecting the effect of ions and assuming only a Maxwellian background. We emphasize that our treatment has been somewhat informal, and is meant to indicate a potential route for future research on the topology and geometry of kinetic plasma waves. The formalism described should also allow for the topological analysis of magnetized kinetic plasmas by appropriately modifying $\mbf{D}_E$.

\section{Conclusion}
We have shown that the rotational symmetry of waves in a cold isotropic plasma induces a natural decomposition into $R$, $L$, and Langmuir waves. The state space for each type of wave globally forms a topological object known as a vector bundle. Langmuir waves have Chern number $C(\zeta_0) = 0$ and are topologically trivial, while $R$ and $L$ waves are topologically nontrivial with $C(\zeta_\pm) = \mp2$. This topological nontriviality has surprising implications, particularly pertaining to the angular momentum. Instead of the ordinary spherical harmonics which describe the orbital angular momentum eigenstates of massive particles, we find that the total angular momentum eigenstates of $R$ and $L$ waves are described by SWSHs. This gives a harmonic decomposition of plasma waves, and in particular, it gives a countable basis for monochromatic waves. SWSHs have appeared in many branches of physics, but this is their first known application in plasma physics. The resulting angular momentum multiplets have a very different structure than that which results for massive particles. In particular, the multiplet structure for $R$ and $L$ waves is far sparser than that of massive particles with spin and also has a lower bound of $j = 1$. This angular momentum multiplet structure reflects the fact that plasma waves do not admit an SAM-OAM decomposition, a somewhat surprising result in light of recent work which has implied such a decomposition exists \cite{Bliokh2022}. We show that the total angular momentum can be decomposed into the helicity and orbital quasi-angular momenta, $\mbf{J}_\parallel$ and $\mbf{J}_\perp$, and that this splitting is induced by the rotational symmetry of $R$ and $L$ waves. It has previously been found that $\mbf{J}_\parallel$ is related to the inverse Faraday effect in plasmas \cite{Bliokh2022}. While this splitting is naturally induced by the symmetry of the waves, $\mbf{J}_\parallel$ and $\mbf{J}_\perp$ do not themselves generate rotational symmetries and are thus not legitimate angular momentum operators. Indeed, these operators do not satisfy angular momentum commutation relations and the components have continuous spectra (in contrast to the discrete spectrum of the components of true angular momentum operators). As a result, we find that most of the theoretical tools pertaining to SAM and OAM operators, such as the Clebsch-Gordan formalism and the Casimir invariance of the square angular momentum operators, do not apply to $\mbf{J}_\parallel$ and $\mbf{J}_\perp$, or to any other attempted splitting of $\mbf{J}$. These are important limitations to account for as further applications of this splitting are explored.

\begin{acknowledgments}
This work is supported by U.S. Department of Energy (DE-AC02-09CH11466).
\end{acknowledgments}

\appendix
\section{Expectation values of plasma operators}\label{appendix:A}
In this appendix we show that the expectation values of $H,\mbf{P},\mbf{J},\mbf{J}_\parallel$, and $\mbf{J_\perp}$ can be calculated in terms of integrals over momentum space or position space. Because each of these operators commutes with the Hamiltonian, the expectation values are all conserved quantities. We use the notation $\mbf{E} \doteq \mbf{E}(\kvec)$, $\mbf{E}' \doteq \mbf{E}(\kvec')$, $\bar{\mbf{E}} \doteq \mbf{E}(-\kvec)$, and similarly for other variables. We also write $\omega_\gamma(\kvec) \doteq \omega_+(\kvec) = \omega_-(\kvec)$.
\begin{prop}[Expectation value of $H$]\label{prop:plasma:H_ev}
\begin{subequations} \begin{align}
    \frac{1}{8\pi}\langle \mbf{E} | H |\mbf{E}\rangle  &= \sum_{h}\frac{1}{8\pi}\int \frac{d^3k}{\omega_h}\mbf{E}_h^*(\mbf{k}) \cdot [H\mbf{E}_h](\mbf{k}) \label{eq:plasma:energy_equiv_1}\\
    &= \sum_{h}\frac{1}{8\pi}\int d^3k\,|\mbf{E}_h|^2 \label{eq:plasma:energy_equiv_2}\\
    &= \sum_h\int d^3x \, \Big[\frac{1}{8\pi}\big(\mathcal{E}^2_h + \mathcal{B}^2_h\big) + \frac{1}{2}mn_0v^2_h\Big] \label{eq:plasma:energy_equiv_3}\\
    &= \int d^3x \, \Big[\frac{1}{8\pi}\big(\mathcal{E}^2 + \mathcal{B}^2\big) + \frac{1}{2}mn_0v^2)\Big] \label{eq:plasma:energy_equiv_4}
\end{align} \end{subequations}
\end{prop}
\begin{proof}
    We can decompose any state into eigenvectors of the helicity $\chi$:
    \begin{equation}
        \mbf{E} = \sum_{h = 0,\pm1} \mbf{E}_h
    \end{equation}
    Since $\chi$ commutes with $H$ and since states with different helicity are orthogonal, it follows that
    \begin{equation}
        \langle\mbf{E}|H|\mbf{E}\rangle = \sum_h\langle \mbf{E}_h|H|\mbf{E}_h\rangle = \sum_{h}\int \frac{d^3k}{\omega_h}\mbf{E}_h^*(\mbf{k}) \cdot [H\mbf{E}_h](\mbf{k}),
    \end{equation}
    establishing Eq. (\ref{eq:plasma:energy_equiv_1}), and $[H\mbf{E}](\kvec) = \omega_h\mbf{E}(\kvec)$ gives Eq. (\ref{eq:plasma:energy_equiv_2}). Analogous statements for the operators $\mbf{J},\mbf{P},\mbf{J}_\parallel$ and $\mbf{J}_\perp$ also hold because these commute with $\chi$. We establish the equivalence of (\ref{eq:plasma:energy_equiv_2}) and (\ref{eq:plasma:energy_equiv_3}) by showing that the summands are equal for each $h$. We have that
    \begin{subequations}\begin{align}
        \boldcal{E}_h(\mbf{x}) &= \frac{1}{2(2\pi)^{3/2}}\int d^3k\, \Big(  \mbf{E}_h(\mbf{k})e^{i(\mbf{k}\cdot \mbf{x}-\omega t)} + c.c.\Big) \\
        \boldcal{B}_h(\mbf{x}) &= \frac{c}{2(2\pi)^{3/2}}\int \frac{d^3k}{\omega_h}\, \Big(  \kvec \times \mbf{E}_h(\mbf{k})e^{i(\mbf{k}\cdot \mbf{x}-\omega t)} + c.c.\Big)
    \end{align}\end{subequations}
    so
    \begin{alignat}{2}
        \int d^3x \, |\boldcal{E}_h|^2 &= \frac{1}{4(2\pi)^3}\int &&d^3x\, d^3k \,d^3k' \Big( \mbf{E}_he^{i(\kvec\cdot \mbf{x}-\omega t)} + \mbf{E}^*_he^{-i(\kvec\cdot \mbf{x}-\omega t)}\Big)\nonumber\\ 
         & && \cdot \Big(  \mbf{E}_h'e^{i(\kvec'\cdot \mbf{x}-\omega' t)} + \mbf{E}'^*_he^{-i(\kvec'\cdot \mbf{x}-\omega' t)}\Big) 
    \end{alignat}
    Using
    \begin{equation}
        \int d^3x \,e^{i\kvec\cdot \mbf{x}} = (2\pi)^3\delta(\mbf{k})
    \end{equation}
    we obtain
    \begin{alignat}{2}
        \frac{1}{8\pi}\int d^3x \, |\boldcal{E}_h|^2 &= \frac{1}{32\pi}\int  d^3k \,\Big( 2 |\mbf{E}_h|^2 &&+\mbf{E}_h\cdot \bar{\mbf{E}}_h e^{-2i\omega t} \nonumber\\
        & &&+ \mbf{E}_h^* \cdot\bar{\mbf{E}}_h^*e^{2i\omega t}\Big) \label{eq:plasma:E_squared_int}
    \end{alignat}
    Using 
    \begin{equation}
        \mbf{v}_h(\kvec) = -i\frac{e}{m\omega_h}\mbf{E}_h(\kvec)
    \end{equation}
    we similarly find
    \begin{alignat}{2}
        \frac{mn_0}{2}\int d^3x \, v^2_h &= \frac{mn_0}{8}\int d^3k \,\Big( 2 |\mbf{v}_h|^2 +\mbf{v}_h\cdot \bar{\mbf{v}}_he^{-2i\omega t} \nonumber\\
        & \qquad\qquad\qquad\qquad\quad\;\; + \mbf{v}^*_h \cdot \bar{\mbf{v}}^*_h e^{2i\omega t}\Big) \\
        &= \frac{1}{32\pi}\int d^3k \,\frac{\omega_p^2}{\omega_h^2}\Big( 2 |\mbf{E}_h|^2 -\mbf{E}_h\cdot \bar{\mbf{E}}_h e^{-2i\omega t} \nonumber\\
        & \qquad\qquad\qquad\qquad\quad\;\;- \mbf{E}_h^* \cdot \bar{\mbf{E}}_h^*e^{2i\omega t}\Big) \label{eq:plasma:v_squared_int}
    \end{alignat}
    
    ($h=0$ case): 
    In the $h=0$ case, $\omega_0 = \omega_p$ and $\boldcal{B} =0$, so from Eqs. (\ref{eq:plasma:E_squared_int}) and (\ref{eq:plasma:v_squared_int}) we have
    \begin{equation}
        \int d^3x \, \Big[\frac{1}{8\pi}\big(\mathcal{E}^2_0 + \mathcal{B}^2_0\big) + \frac{1}{2}mn_0v^2\Big] = \frac{1}{8\pi}\int d^3k \, |\mbf{E}_0|^2
    \end{equation}
    giving the desired result.

    ($h=\pm1$ case):
    In the EM case, $\omega^2_{\pm} = \omega_p^2 + |\kvec|^2c^2$ and $\mbf{k}\cdot \mbf{E}_{\pm} = 0$. By replacing $\boldcal{E}$ with $\boldcal{B}$ in Eq. (\ref{eq:plasma:E_squared_int}), and using the relationship
    \begin{equation}
        \mbf{B}_{\pm} = \frac{c|\kvec|}{\omega_\gamma}\khat \times \mbf{E}_{\pm}
    \end{equation}
    we obtain
    \begin{alignat}{2}
        \frac{1}{8\pi}\int d^3x \, \mathcal{B}^2_{\pm} &= \frac{1}{32\pi}\int  d^3k \frac{c^2|\kvec|^2}{\omega^2_\gamma}\,\Big( 2 |\mbf{E}_\pm|^2 &&-\mbf{E}_\pm \cdot \bar{\mbf{E}}_\pm e^{-2i\omega t} \nonumber\\
        & &&+ \mbf{E}_\pm^* \cdot \bar{\mbf{E}}_\pm^*e^{2i\omega t}\Big) \label{eq:plasma:B_squared_int}
    \end{alignat}
    This equation combines with Eqs. (\ref{eq:plasma:E_squared_int}) and (\ref{eq:plasma:v_squared_int}) to give
    \begin{equation}
        \int d^3x \, \Big[\frac{1}{8\pi}\big(\mathcal{E}^2_{\pm} + \mathcal{B}^2_{\pm}) + \frac{1}{2}mn_0v^2_{\pm}\Big] = \frac{1}{8\pi}\int d^3k \, |\mbf{E}_{\pm}|^2,
    \end{equation}
    completing the EM case and establishing the equivalence of (\ref{eq:plasma:energy_equiv_2}) and (\ref{eq:plasma:energy_equiv_3}). Lastly, we note that equation (\ref{eq:plasma:energy_equiv_4}) is established if we show that the cross terms in the inner products vanish, namely, if
    \begin{equation}\label{eq:plasma:energy_cross_terms}
        \int d^3x \, \Big[\frac{1}{8\pi}\big(\boldcal{E}_h \cdot\boldcal{E}_{h'} + \boldcal{B}_h\cdot\boldcal{B}_{h'}\big) + \frac{1}{2}mn_0\mbf{v}_h\cdot\mbf{v}_{h'}\Big]  = 0
    \end{equation}
    for $h\neq h'$. In general,
    \begin{equation}
        \int d^3x\,\boldcal{E}_h \cdot\boldcal{E}_{h'} = \frac{1}{4}\int d^3k\big[\mbf{E}_h\cdot\mbf{E}^*_{h'}e^{-i(\omega_h-\omega_{h'})} + \mbf{E}_h\cdot \bar{\mbf{E}}_{h'}e^{-i(\omega_h+\omega_{h'})} + c.c.].
    \end{equation}
    The first term vanishes since $\eh \cdot \uv{e}^*_{h'} = 0$ for $h \neq h'$. If exactly one of $h$ or $h'$ is $0$, then the second dot product clearly vanishes. If $h = \pm1$ and $h'=\mp1$, then since $\mbf{E}_\mp(-\kvec)$ is proportional to $\mbf{E}_\pm(\kvec)$,
    \begin{equation}
        \mbf{E}_{\pm}(\kvec) \cdot \mbf{E}_{\mp}(-\kvec) = c\mbf{E}_{\pm}(\kvec) \cdot \mbf{E}_{\pm}(\kvec) = 0.
    \end{equation}
    Thus
    \begin{equation}
        \int d^3x\,\boldcal{E}_h \cdot\boldcal{E}_{h'} = 0.
    \end{equation}
    Similar calculations show that the other two terms on the lhs of (\ref{eq:plasma:energy_cross_terms}) vanish, establishing (\ref{eq:plasma:energy_equiv_4}).
\end{proof}
\begin{prop}[Expectation value of $\mbf{P}$]\label{prop:plasma:P_ev}
\begin{subequations} \begin{align}
    \frac{1}{8\pi}\langle \mbf{E} | \mbf{P}|\mbf{E}\rangle &= \sum_h\frac{1}{8\pi} \int \frac{d^3k}{\omega_h}\mbf{E}^*_h(\mbf{k}) \cdot [\mbf{P}\mbf{E}_h](\mbf{k}) \label{eq:plasma:momentum_equiv_1}\\
    &= \sum_h\frac{1}{8\pi} \int d^3k\frac{\kvec}{\omega_h}|\mbf{E}_h|^2 \label{eq:plasma:momentum_equiv_2}\\
    &= \sum_h\int d^3x \, \Big[\frac{1}{4\pi c}\boldcal{E}_h \times \boldcal{B}_h + mn_{1,h}\mbf{v}_h\Big] \label{eq:plasma:momentum_equiv_3} \\
    &= \int d^3x \, \Big[\frac{1}{4\pi c}\boldcal{E} \times \boldcal{B} + mn_{1,h}\mbf{v}\Big] \label{eq:plasma:momentum_equiv_4}
\end{align} \end{subequations}
\end{prop}
\begin{proof}
From $\mbf{P}\mbf{E}(\kvec) = \mbf{k}\mbf{E}(\kvec)$ we obtain
    \begin{equation}
        \langle \mbf{E} | \mbf{P} |\mbf{E} \rangle = \sum_h\int d^3k \,\frac{\kvec_h}{\omega_h}|\mbf{E}_h|^2.
    \end{equation}
    We now show that the summands in Eqs. (\ref{eq:plasma:momentum_equiv_2}) and (\ref{eq:plasma:momentum_equiv_3}) are equal for each $h$. 

    ($h = \pm 1$ case): If $h = \pm1$, then $n_{1,h} = 0$, so the only nonvanishing term on the rhs of Eq. (\ref{eq:plasma:momentum_equiv_3}) is the EM term. We have
    \begin{align}
        \int d^3x\, \boldcal{E}_{h} \times \boldcal{B}_h = \frac{1}{4(2\pi)^3}\int  d^3x \,d^3k \, d^3k' \,\frac{c|\kvec'|}{\omega'_h}\Big(  \mbf{E}_he^{i(\kvec\cdot \mbf{x}-\omega_h t)} + \mbf{E}^*_he^{-i(\kvec\cdot \mbf{x}-\omega_h t)}&\Big)\nonumber \\
        \times  \Big(  \khat' \times\mbf{E}'_he^{i(\kvec'\cdot \mbf{x}-\omega_h't)} + \khat' \times\mbf{E}'^*_he^{-i(\kvec'\cdot \mbf{x}-\omega_h' t)}&\Big) \label{eq:plasma:mom_cons_1}
    \end{align}
    Note that
    \begin{align}
        I = \frac{1}{(2\pi)^3}\int d^3x \,d^3k \, &d^3k' \, \frac{c|\kvec'|}{\omega'_h} \mbf{E}_h \times \big(\khat' \times \mbf{E}'_h\big)e^{i\mbf{x}\cdot(\kvec + \kvec')-it(\omega +\omega')} \nonumber\\
        &= -\int d^3k \, \frac{c|\kvec|}{\omega'_h} \mbf{E}_h \times \big(\khat \times \mbf{E}'_h\big)e^{-2i\omega_h t}
    \end{align}
    By substituting $\kvec \rightarrow -\kvec$ and noting that
    \begin{equation}
        \mbf{E}_{\pm}(-\kvec) \times\big(\khat \times\mbf{E}_{\pm}(\kvec)\big) = \mbf{E}_{\pm}(\kvec) \times\big(\khat \times\mbf{E}_{\pm}(-\kvec)\big),
    \end{equation}
    we see that $I = -I$ so that
    \begin{equation}
        I = 0.
    \end{equation}
    Similarly, the cross term in Eq. (\ref{eq:plasma:mom_cons_1}) involving $\mbf{E}^*(\kvec)$ and $\mbf{E}^*(\kvec')$ also vanishes. The remaining two terms give
    \begin{alignat}{2}
        \frac{1}{4\pi c}\int d^3x\, \boldcal{E}_h \times \boldcal{B}_h &= \frac{1}{8\pi} \int d^3k \frac{\kvec}{\omega_h}|\mbf{E}_h|^2 
    \end{alignat}
    for $h = \pm1$.

    (h = 0 case): For Langmuir waves we have $\boldcal{B}_0 = 0$ and in Fourier space
    \begin{subequations}\begin{gather}
        n_{1,0} = -\frac{i}{4\pi e}\kvec \cdot \mbf{E}_0 \label{eq:plasma:n1_ES}\\
        \mbf{v}_h = -\frac{ie}{m\omega_h}\mbf{E}_h \label{eq:plasma:v_ES}\\
        (\mbf{E}_0\cdot \kvec)\mbf{E}_0^* = \kvec |\mbf{E}_0|^2.
    \end{gather} \end{subequations}
    The only nonvanishing term on the rhs of Eq. (\ref{eq:plasma:momentum_equiv_3}) is
    \begin{subequations}\begin{alignat}{1}
        \int d^3x \,mn_{1,0}\mbf{v}_0 &= -\frac{1}{16\pi(2\pi)^3}\int d^3x \, d^3k \, d^3k' \frac{1}{\omega_p}\big[\kvec\cdot(\mbf{E}_0e^{i(\kvec\cdot \mbf{x}-\omega_p t)} -\mbf{E}_0^*e^{-i(\kvec\cdot \mbf{x}-\omega_p t)})] \nonumber\\
        & \qquad\qquad\qquad\qquad\qquad\quad\qquad\times \big(\mbf{E}'_0 e^{i(\kvec'\cdot \mbf{x}-\omega_p t)} -\mbf{E}'^*_0e^{-i(\kvec'\cdot \mbf{x}-\omega_p t)} \big) \\
        &= \frac{1}{8\pi}\int d^3k \, \frac{\kvec}{\omega_p}|\mbf{E}_0|^2 + \Big(e^{-2i\omega_pt}\int\frac{d^3k}{\omega_p}(\kvec\cdot\mbf{E}_0)\bar{\mbf{E}}_0 + c.c.\Big).
    \end{alignat}\end{subequations}
    The last vanishes is zero because by writing $\mbf{E}_0(\kvec) = \alpha(\kvec)\khat$ and taking $\kvec \rightarrow -\kvec$:
    \begin{subequations}\begin{align}
        \int d^3k(\kvec\cdot\mbf{E}_0)\bar{\mbf{E}}_0 &= \int d^3k \,\alpha(\kvec)\alpha(-\kvec)\khat \nonumber\\
        &= -\int d^3k \alpha(-\kvec)\alpha(\kvec)\khat.
    \end{align}\end{subequations}
    Therefore 
    \begin{equation}
        \int d^3x \,mn_{1,0}\mbf{v}_0 
        = \frac{1}{8\pi}\int d^3k \, \frac{\kvec}{\omega_p}|\mbf{E}_0|^2,
    \end{equation}
    establishing the equality of Eqs. (\ref{eq:plasma:momentum_equiv_2}) and (\ref{eq:plasma:momentum_equiv_3}).

    To show the equivalence of Eqs. (\ref{eq:plasma:momentum_equiv_3}) and (\ref{eq:plasma:momentum_equiv_4}) it is sufficient to show that the helicity cross terms in (\ref{eq:plasma:momentum_equiv_4}) vanish, namely that, 
    \begin{equation}
        \int d^3x \, \Big[\frac{1}{4\pi c}\boldcal{E}_h \times \boldcal{B}_{h'} + mn_{1,h}\mbf{v}_{h'
        }\Big] = 0 \label{eq:plasma:momentum_cross_terms}
    \end{equation}
    for $h \neq h'$. For $h = \pm1$ and $h' = \mp1$:
    \begin{align}
         \int d^3x\, \boldcal{E}_{h} \times \boldcal{B}_{h'} = \frac{1}{4(2\pi)^3}\int  d^3x \,d^3k \, d^3k' \,\frac{c|\kvec'|}{\omega'_\gamma}\Big(  \mbf{E}_{\pm}e^{i(\kvec\cdot \mbf{x}-\omega_\gamma t)} + \mbf{E}^*_{\pm}e^{-i(\kvec\cdot \mbf{x}-\omega_\gamma t)}&\Big)\nonumber \\
        \times  \Big(  \khat' \times\mbf{E}'_{\mp}e^{i(\kvec'\cdot \mbf{x}-\omega'_\gamma t)} + \khat' \times\mbf{E}'^*_{\mp}e^{-i(\kvec'\cdot \mbf{x}-\omega'_\gamma t)}&\Big). \label{eq:plasma:momentum_EM_cross_terms}
    \end{align}
    Then
    \begin{subequations}\begin{align}
        \frac{1}{(2\pi)^3}\int d^3x \,d^3k \, &d^3k' \, \frac{c|\kvec'|}{\omega'_\gamma} \mbf{E}_{\pm} \times \big(\khat' \times \mbf{E}'_{\mp}\big)e^{i\mbf{x}\cdot(\kvec + \kvec')-it(\omega_\gamma +\omega'_\gamma)} \\
        &= -\int d^3k \, \frac{c|\kvec|}{\omega_\gamma} \mbf{E}_{\pm} \times \big(\khat \times \bar{\mbf{E}}_{\mp}\big)e^{-2i\omega_\gamma t} \\
        &= \mp i\int d^3k \, \frac{c\kvec}{\omega_\gamma} [\mbf{E}_{\pm}(\kvec) \cdot \bar{\mbf{E}}_{\mp}]e^{-2i\omega_\gamma t} \\
        &= 0.
    \end{align}\end{subequations}
    The last line follows because $\mbf{E}_{\mp}(-\mbf{k}) = C(\mbf{k})\mbf{E}_{\pm}(\kvec)$ for some scalar function $C$ and $\mbf{E}_{\pm}(\kvec) \cdot \mbf{E}_{\pm}(\kvec) = 0$. The term involving $\mbf{E}_{\pm}^*$ and $\mbf{E}_{\mp}'^*$ in Eq. (\ref{eq:plasma:momentum_EM_cross_terms}) similarly vanishes. The term involving $\mbf{E}_{\pm}$ and $\mbf{E}_{\mp}^*$ (and the term involving their conjugates) also vanishes:
    \begin{subequations}\begin{align}
        \frac{1}{(2\pi)^3}\int d^3x \,d^3k \, &d^3k' \, \frac{c|\kvec'|}{\omega'} \mbf{E}_{\pm} \times \big(\khat' \times \mbf{E}'^*_{\mp}\big)e^{i\mbf{x}\cdot(\kvec + \kvec')-it(\omega_\gamma +\omega'_\gamma)} \\
        &= -\int d^3k \, \frac{c|\kvec|}{\omega_\gamma} \mbf{E}_{\pm} \times \big(\khat \times \mbf{E}_{\mp}^*\big)e^{-2i\omega_\gamma t} \\
        &= -\int d^3k \, \frac{c\kvec}{\omega_\gamma} (\mbf{E}_{\pm} \cdot \mbf{E}_{\mp}^*)e^{-2i\omega_\gamma t} \\
        &= 0.
    \end{align}\end{subequations}
    Thus Eq. (\ref{eq:plasma:momentum_cross_terms}) holds for $h = \pm 1$ and $h' = \mp1$. The $h = \pm1 $, $h' = 0$ case is trivial since $n_{1,\pm} = 0$ and $\mathcal{B}_0 = 0$. The final case, $h = 0$ and $h' = \pm 1$, follows because
    \begin{subequations}\begin{align}
        \int d^3x \, mn_{1,0}v_{\pm} &= -\frac{1}{16\pi}\int \frac{d^3k}{\omega_\gamma}\Big[(\kvec\cdot\mbf{E}_0)\bar{\mbf{E}}_{\pm}e^{-it(\omega_\gamma + \omega_p)} \nonumber\\
        & \qquad\qquad\qquad-(\kvec\cdot\mbf{E}_0)\bar{\mbf{E}}^*_{\pm}e^{it(\omega_{\gamma} - \omega_p)}   + c.c.\Big] \\
        &=-\int d^3x \, \frac{1}{4\pi c}\boldcal{E}_0 \times \boldcal{B}_\pm.
    \end{align}\end{subequations}
\end{proof}
\begin{prop}[Expectation value of $\mbf{J}$]\label{prop:plasma:J_ev}
\begin{subequations} \begin{align}
        \frac{1}{8\pi}\langle \mbf{E} | \mbf{J}|\mbf{E}\rangle &= \sum_h\frac{1}{8\pi}\int \frac{d^3k}{\omega_h}\mbf{E}^*_h(\mbf{k}) \cdot [\mbf{J}\mbf{E}_h](\mbf{k}) \label{eq:plasma:AM_equiv_1}\\
        &= -i\sum_{h}\frac{1}{8\pi}\int \frac{d^3k}{\omega_h}\Big(\sum_{n}E^*_{h,n}\cdot (\kvec \times \nabla_{\kvec})E_{h,n} + \mbf{E}^*_h \times \mbf{E}_h\Big) \label{eq:plasma:AM_equiv_2}\\
        &= \sum_h\int d^3x \, \Big[\frac{1}{4\pi c}\mbf{x} \times \big(\boldcal{E}_h \times \boldcal{B}_h\big) + mn_{1,h}\mbf{x} \times \mbf{v}_h\Big] \label{eq:plasma:AM_equiv_3} \\
        &= \int d^3x \, \Big[\frac{1}{4\pi c}\mbf{x} \times \big(\boldcal{E} \times \boldcal{B}\big) + mn_1\mbf{x} \times \mbf{v}\Big] \label{eq:plasma:AM_equiv_4}
\end{align} \end{subequations}
\end{prop}
\begin{proof}
    It was shown in Ref. \cite{PalmerducaQin_connection}, Lemma 1 that $\mbf{J}_\parallel$ and $\mbf{J}_\perp$ can be expressed as
    \begin{subequations}\begin{align}
        \mbf{J}_\parallel = \mathcal{P}_h \circ\mbf{S} \\
        \mbf{J}_\perp = \mathcal{P}_h \circ{\mbf{L}}
    \end{align}\end{subequations}
    where $\mbf{S}$ and $\mbf{L}$ are the usual spin and orbital angular momentum operators
    \begin{subequations}\begin{gather}
        (S_a)_{bc} = -i\epsilon_{abc} \\
        \mbf{L} = -i\khat \times \nabla_{\kvec}
    \end{gather}\end{subequations}
    and $\mathcal{P}:\pspace\times \Comp^3 \rightarrow\zeta_h$ is the projection onto the helicity $h$ states. For a component $J_a$ of the angular momentum, we generally have
    \begin{subequations}
    \begin{align}
        \langle \mbf{E}|J_a|\mbf{E} \rangle &= \sum_h \int \frac{d^3 k}{\omega} \mbf{E}^*_h(\kvec) \cdot [J_a\mbf{E}_h](\kvec) \\ 
        &= \sum_h\int \frac{d^3 k}{\omega} \Big(\mbf{E}^*_h\cdot (\mathcal{P}_h \circ L_a)\mbf{E}_h + \mbf{E}^*_h\cdot (\mathcal{P}_h \circ S_a)\mbf{E}_h \Big) \\
        &= \sum_h\int \frac{d^3 k}{\omega} \Big(\mbf{E}^*_h\cdot L_a\mbf{E}_h + \mbf{E}^*_h\cdot S_a\mbf{E}_h\Big) \\
        &= -i\int \frac{d^3 k}{\omega} \Big(\mbf{E}^*_h\cdot (\kvec \times \nabla_{\kvec})_a\mbf{E}_h + E^*_{h,b}\epsilon_{abc}E_{h,c} \Big) \\
        &= -i\int \frac{d^3 k}{\omega} \Big(\sum_{n}E^*_{h,n} (\kvec \times \nabla_{\kvec})_aE_{h,n} + (\mbf{E}^*_h \times \mbf{E}_h)_a\Big),
    \end{align}
    \end{subequations}
    establishing Eqs. (\ref{eq:plasma:AM_equiv_1}) and (\ref{eq:plasma:AM_equiv_2}). We show that the equivalence of (\ref{eq:plasma:AM_equiv_2}) and (\ref{eq:plasma:AM_equiv_3}) by showing the integrands are equal for each $h$.
    
    ($h=\pm1$ case): In the EM case, $n_{1,\pm} = 0$, so the claim is that
    \begin{equation}
        \frac{1}{4\pi c}\int d^3x \, \mbf{x} \times \big(\boldcal{E}_\pm \times \boldcal{B}_\pm\big) =-i\int \frac{d^3 k}{\omega} \Big(\sum_{n}E^*_{\pm,n}\cdot (\kvec \times \nabla_{\kvec})E_{\pm,n} + \mbf{E}^*_\pm \times \mbf{E}_\pm\Big)
    \end{equation}
    One can prove this by similar techniques used for Propositions \ref{prop:plasma:H_ev} and \ref{prop:plasma:P_ev}. However, we omit the proof since a detailed derivation of this result is given by Li in the Appendix of Ref. \cite{Li2009}. While Li was concerned with the case of EM waves in the vacuum, the derivation only uses the relationship
    \begin{equation}
        \boldcal{B}(\mbf{x}) = \frac{1}{2(2\pi)^{3/2}}\int \frac{d^3k}{\omega}\, \Big(  \kvec \times \mbf{E}e^{i(\mbf{k}\cdot \mbf{x}-\omega t)} + c.c.\Big)
    \end{equation}
    and is equally valid in the vacuum ($\omega = |\kvec|$) and plasma $\big(\omega = \sqrt{\omega_p^2 + c^2|\kvec|^2}\,\big)$ cases.
    
    (h=0 case): For Langmuir waves, $\mbf{E}^*_0 \times \mbf{E}_0 = 0$, $\boldcal{B}_0 = 0$, and we can write $\mbf{E}(\kvec) = \alpha(\kvec)\khat$. We thus wish to show that
    \begin{subequations}\begin{align}
        \int d^3x m_e n_{1,0}\mbf{x} \times\mbf{v}_0 &= -\frac{i}{8\pi}\int \frac{d^3k}{\omega_p} \mbf{E}_0^*\cdot (\mbf{k} \times \nabla_{\mbf{k}})\mbf{E}_0 \\
        &= -\frac{i}{8\pi}\int \frac{d^3k}{\omega_p} \alpha^*(\mbf{k} \times \nabla_{\mbf{k}})\alpha \label{eq:plasma:AM_ES_rhs}.
    \end{align}\end{subequations}
    Starting with the lhs, we have
    \begin{subequations}\begin{alignat}{2}
        \int d^3x m_e n_{1,0}\mbf{x} \times\mbf{v}_0 &= \frac{m_e}{4(2\pi)^3}\int d^3x \, d^3k \, d^3 k' \, \mbf{x}\times&&\big[(n_{1,0}(\kvec')e^{i(\kvec'\cdot \mbf{x} - \omega_p t)} + c.c.) \\
         & && \cdot (\mbf{v}_{0}(\kvec)e^{i(\kvec\cdot \mbf{x} - \omega_p t)} + c.c.)\big].
    \end{alignat}\end{subequations}
    Using Eqs. (\ref{eq:plasma:n1_ES}) and (\ref{eq:plasma:v_ES}) we get
    \begin{subequations}\begin{gather}
        \int d^3x m_e n_{1,0}\mbf{x} \times\mbf{v}_0 = \frac{1}{8\pi}\big(\re(I_1) + \re(I_2)\big) \label{eq:plasma:ES_AM_I_1_I_2}\\
        I_1 = \frac{1}{(2\pi)^3}\int d^3x \, d^3k \, d^3k' \, \frac{d^3 k}{\omega_p}(\kvec'\cdot \mbf{E}'_0)\mbf{E}_0^* \times \mbf{x} e^{i\mbf{x}\cdot(\kvec-\kvec')} \\
        I_2 = -\frac{1}{(2\pi)^3}\int d^3x \, d^3k \, d^3k' \, \frac{d^3 k}{\omega_p}(\kvec'\cdot \mbf{E}'_0)e^{-2i\omega_p t}\mbf{E}_0 \times \mbf{x} e^{i\mbf{x}\cdot(\kvec+\kvec')}.
    \end{gather}\end{subequations}
    Using 
    \begin{equation}
        \int d^3x \,\mbf{x}e^{i\mbf{x}\cdot(\kvec - \kvec')} = -(2\pi)^3 i \nabla_{\kvec'}\delta(\kvec - \kvec')
    \end{equation}
    we have
    \begin{subequations}\begin{align}
        I_1 &= -i \int \frac{d^3k}{\omega_p}d^3k'\, (\kvec' \cdot \mbf{E}'_0)\nabla_{\kvec'}\times (\mbf{E}^*_0\delta(\kvec - \kvec')) \\
        &= i \int \frac{d^3k}{\omega_p}\nabla_{\kvec}(\kvec \cdot \mbf{E}_0)\times\mbf{E}^*_0 \\
        &= -i\int\frac{d^3k}{\omega_p}\alpha^*\kvec \times\nabla_{\kvec}\alpha \label{eq:plasma:ES_I_1}\\
        &= \langle \mbf{E}_0 | \mbf{J} | \mbf{E}_0\rangle \in \Real
    \end{align}\end{subequations}
Similar algebra shows that
    \begin{subequations}\begin{align}
        I_2 &=  -ie^{-2i\omega_pt}\int\frac{d^3k}{\omega_p}\alpha(\kvec)\kvec \times(\nabla_{\kvec}\alpha)(-\kvec) \\
        &=  i e^{-2i\omega_pt} \int\frac{d^3k}{\omega_p}\alpha(-\kvec)\kvec \times (\nabla_{\kvec}\alpha)(\kvec) \\
        &= ie^{-2i\omega_pt} \Big(\int\frac{d^3k}{\omega_p}\alpha(\kvec)\kvec \times(\nabla_{\kvec}\alpha)(-\kvec) \\
        &\qquad\qquad\qquad-\int d^3k \, \nabla_{\kvec} \times\Big[\frac{\alpha(\kvec) \alpha(-\kvec)\kvec}{\omega_p} \Big] \Big)\\
        &= -I_2
    \end{align}\end{subequations}
    so that
    \begin{equation}\label{eq:plasma:ES_I_2}
        I_2 = 0.
    \end{equation}
    Combining Eqs. (\ref{eq:plasma:ES_AM_I_1_I_2}), (\ref{eq:plasma:ES_I_1}), and (\ref{eq:plasma:ES_I_2}) shows that
    \begin{equation}
        \int d^3x m_e n_{1,0}\mbf{x} \times\mbf{v}_0 = -\frac{i}{8\pi}\int \frac{d^3k}{\omega_p} \alpha^*(\mbf{k} \times \nabla_{\mbf{k}})\alpha,
    \end{equation}
    establishing the equivalence of (\ref{eq:plasma:AM_equiv_2}) and (\ref{eq:plasma:AM_equiv_3}).

    We verify that (\ref{eq:plasma:AM_equiv_3}) and (\ref{eq:plasma:AM_equiv_4}) are equivalent by checking that the helicity cross terms ($h \neq h'$) vanish:
    \begin{equation}\label{eq:plasma:AM_cross_terms}
    \int d^3x \, \Big[\frac{1}{4\pi c}\mbf{x} \times \big(\boldcal{E}_h \times \boldcal{B}_{h'}\big) + mn_{1,h}\mbf{x} \times \mbf{v}_{h'}\Big] = 0.
    \end{equation}
    This is trivial for $h = \pm1$ and $h' = 0$ since $\boldcal{B}_0 = 0$ and $n_{1,\pm} = 0$. For the $h = 0$  and $h' = \pm1$, it is straight-forward to show that
    \begin{subequations}\begin{alignat}{2}
        \int d^3x \,&m n_{1,0}\mbf{x}\times\mbf{v}_{\pm} \\
        &= \int d^3k \frac{i}{4\pi\omega_\gamma}\, \Big[-e^{it(\omega_\pm - \omega_p)}\nabla_{\kvec}[\kvec\cdot\mbf{E}_0]\times\mbf{E}^*_\pm \nonumber\\
        & \qquad\qquad\qquad\,+e^{-it(\bar{\omega}_\pm + \omega_p)}\nabla_{\kvec}[\kvec\cdot\mbf{E}_0]\times\bar{\mbf{E}}_\pm\Big] + c.c. \\
        &= -\frac{1}{4\pi c}\int d^3x \mbf{x}\times (\boldcal{E}_0\times \boldcal{B}_\pm)
    \end{alignat}\end{subequations}
    so that Eq. (\ref{eq:plasma:AM_cross_terms}) holds. For $h = \pm 1$ and $h = \mp1$ the second term in Eq. (\ref{eq:plasma:AM_cross_terms}) clearly vanishes. The remaining EM term can be split into two parts:
    \begin{subequations}\begin{gather}
        \int d^3x \,\mbf{x}\times(\boldcal{E}_\pm \times \boldcal{B}_\mp) = I_3 + I_4 \\
        I_3 = \int d^3x \, d^3k \,d^3k' \,\mbf{x}\times(\mbf{E}_\pm e^{i(\mbf{k}\cdot\mbf{x} - \omega_\gamma t)} \times {\mbf{B}'^*_\mp} e^{-i(\mbf{k}'\cdot\mbf{x} -\omega_\gamma' t)}) + c.c. \\
        I_4 = \int d^3x \, d^3k \,d^3k' \,\mbf{x}\times(\mbf{E}_\pm e^{i(\mbf{k}\cdot\mbf{x} - \omega_\gamma t)} \times {\mbf{B}'_\mp} e^{i(\mbf{k}'\cdot\mbf{x} - \omega_\gamma' t)}) + c.c. .
    \end{gather}\end{subequations}
    Standard manipulations, along with the fact that $\mbf{E}_\mp^* \times \mbf{E}_\pm = 0$, give
    \begin{align}
        I_3 &= \pm\int d^3k \frac{c |\kvec|}{\omega_\gamma}\nabla\times(\mbf{E}_\mp^* \times \mbf{E}_\pm) = 0.
    \end{align}
    Similarly, since $\bar{\mbf{E}}_\mp \times \mbf{E}_\pm = 0$, we have
    \begin{equation}
        I_4 = \pm \int d^3k\, \frac{c|\mbf{k}|}{2\omega_\gamma} \nabla \times (\bar{\mbf{E}}_{\mp} \times \mbf{E}_\pm) + c.c. = 0.
    \end{equation}
    Thus
    \begin{equation}
        \int d^3x \,\mbf{x}\times(\boldcal{E}_\pm \times \boldcal{B}_\mp) = 0,
    \end{equation}
    completing the verification.
\end{proof}
\begin{prop}[Expectation value of $\mbf{J}_\parallel$]\label{prop:plasma:J_par_ev}
\begin{subequations} \begin{align}
        \frac{1}{8\pi}\langle \mbf{E} | \mbf{J}_\parallel|\mbf{E}\rangle &= \sum_h\frac{1}{8\pi}\int \frac{d^3k}{\omega_h}\mbf{E}^*_h(\mbf{k}) \cdot [\mbf{J}_\parallel \mbf{E}_h](\mbf{k}) \label{eq:plasma:par_equiv_1}\\
        &= -i\sum_{h}\frac{1}{8\pi}\int \frac{d^3k}{\omega_h}\mbf{E}^*_h \times \mbf{E}_h\label{eq:plasma:par_equiv_2}\\
        &= \sum_h \frac{1}{4\pi c}\int d^3x \, \boldcal{E}^\perp_h \times \boldcal{A}^\perp_h \label{eq:plasma:par_equiv_3} \\
        &= \frac{1}{4\pi c}\int d^3x \, \boldcal{E}^{\perp} \times \boldcal{A}^{\perp} \label{eq:plasma:par_equiv_4}
\end{align} \end{subequations}
\end{prop}
\begin{proof}
    Eqs. (\ref{eq:plasma:par_equiv_1}) and (\ref{eq:plasma:par_equiv_2}) follow from considering only the $\mbf{J}_\parallel = \mathcal{P}_h\circ\mbf{S}$ term in Proposition \ref{prop:plasma:J_ev}. Since $\mbf{E} = \mbf{E}_0 + \mbf{E}^\perp$ and since $\mbf{E}_0^\perp \times \mbf{E}_0 = \mbf{E}_0^*\times\mbf{E}_0 = 0$, the sum can be taken only over $h = \pm1$, or equivalently, $\mbf{E}$ can be replaced by $\mbf{E}^\perp$. Eq. (\ref{eq:plasma:par_equiv_3}) follow from $\boldcal{A}^\perp = c^{-1}\partial_t\boldcal{E}^\perp$. Lastly, we prove the equivalence of (\ref{eq:plasma:par_equiv_3}) and (\ref{eq:plasma:par_equiv_4}). Since $\boldcal{E}^\perp_0 = \boldcal{A}^\perp_0 = 0$, it is sufficient to show that 
    \begin{equation}
        \int d^3x \boldcal{E}^\perp_{\pm} \times \boldcal{A}^\perp_{\mp} = 0.
    \end{equation}
    This follows because
    \begin{align}
        \int d^3x \, \boldcal{E}^\perp_{\pm} \times \boldcal{A}^\perp_{\mp} = \int d^3k \, \frac{1}{\omega_\gamma}\big[&\mbf{E}_\pm \times\mbf{E}_\mp(-\kvec) e^{-2it\omega_{\pm}(\kvec)} \nonumber\\
        &+ \mbf{E}_\pm(\kvec) \times\mbf{E}^*_\mp(\kvec) + c.c.\big].
    \end{align}
    The first term on the rhs vanishes since $\mbf{E}_\mp(-\kvec) = C_1(\kvec)\mbf{E}_\pm(\kvec)$ and the second vanishes since $\mbf{E}_{\mp}^*(\kvec) = C_2(\kvec)\mbf{E}_\pm(\kvec)$ for some scalar functions $C_1$ and $C_2$.
\end{proof}
\begin{prop}[Expectation value of $\mbf{J}_\perp$]\label{prop:plasma:J_perp_ev}
\begin{subequations} \begin{align}
        \frac{1}{8\pi}\langle \mbf{E} | \mbf{J}_\perp|\mbf{E}\rangle &= \sum_h\frac{1}{8\pi}\int \frac{d^3k}{\omega_h}\mbf{E}^*_h(\mbf{k}) \cdot [\mbf{J}_\perp \mbf{E}_h](\mbf{k}) \label{eq:plasma:perp_equiv_1}\\
        &= -i\sum_{h,n}\frac{1}{8\pi}\int \frac{d^3k}{\omega_h}E^*_{h,n} (\kvec \times \nabla_{\kvec})E_{h,n}\label{eq:plasma:perp_equiv_2}\\
        &= \sum_{h}\int d^3x \Big[ \, \frac{1}{4\pi c}\sum_n \mathcal{E}^{\perp}_{h,n} (\mbf{x}\times \nabla) \mathcal{A}^\perp_{h,n} + mn_{1,h}\mbf{x} \times \mbf{v}_h\Big] \label{eq:plasma:perp_equiv_3}\\
        &= \int d^3x \Big[ \, \frac{1}{4\pi c}\sum_n \mathcal{E}^{\perp}_{n} (\mbf{x}\times \nabla) \mathcal{A}^\perp_{n} + mn_{1}\mbf{x} \times \mbf{v}\nonumber\\
        & \qquad\qquad\qquad -\frac{e}{c}n_{1}\mbf{x} \times \boldcal{A}^\perp \Big]\label{eq:plasma:perp_equiv_4}
\end{align} \end{subequations}
\end{prop}
\begin{proof}
Ref. \cite{Cohen1989} (pp. 45-47) gives the decomposition
\begin{align}
    \frac{1}{4\pi c}\int d^3x \, \mbf{x}\times(\boldcal{E} \times\boldcal{B}) = \int d^3x \Big\{ \sum_n &\mathcal{E}_n^\perp (\mbf{x}\times \nabla)\mathcal{A}^\perp_n + \boldcal{E}^\perp \times \boldcal{A}^\perp \nonumber\\
    &-en_1\mbf{x}\times \boldcal{A}^\perp\Big\}.
\end{align}
The equivalence of (\ref{eq:plasma:perp_equiv_1}), (\ref{eq:plasma:perp_equiv_2}), and (\ref{eq:plasma:perp_equiv_4}) then follow directly the corresponding equations for $\mbf{J}$ (Proposition \ref{prop:plasma:J_ev}) and $\mbf{J}_\parallel$ (Proposition \ref{prop:plasma:J_par_ev}) by using $\mbf{J}_\perp = \mbf{J} - \mbf{J}_\parallel$. Those Propositions also directly imply that
\begin{align}
    \frac{1}{8\pi}\langle \mbf{E} | \mbf{J}_\perp|\mbf{E}\rangle = 
    &= \sum_{h}\int d^3x \Big[ \, \frac{1}{4\pi c}\sum_n \mathcal{E}^{\perp}_{h,n} (\mbf{x}\times \nabla) \mathcal{A}^\perp_{h,n} + mn_{1,h}\mbf{x} \times \mbf{v}_h \nonumber\\
    &\qquad \qquad \qquad -en_{1,h}\mbf{x}\times \boldcal{A}^\perp_h]. 
\end{align}
Eq. (\ref{eq:plasma:perp_equiv_3}) then follows because $n_{1,h}\boldcal{A}^\perp_h = 0$ for all $h$.
\end{proof}

\bibliography{plasma_topology}
\end{document}